\documentclass[%
 aip,
 jmp,%
 amsmath,amssymb,
 reprint,%
]{revtex4-2}

\usepackage{graphicx}% Include figure files
\usepackage{dcolumn}% Align table columns on decimal point
\usepackage{bm}% bold math
\usepackage{caption}
\usepackage{subcaption}
\begin{document}

\preprint{AIP/123-QED}

\title{From Brittle to Ductile and Back: Reentrant Fracture Transition in Disordered Two-Phase Solids}% Force line breaks with \\

\author{Subrat Senapati}
\email{subrat.senapati52@gmail.com}
\affiliation{Department of Applied Mechanics, IIT Madras, Chennai 600036, India}%Lines break automatically or can be forced with \\
\author{Anuradha Banerjee}%
\email{anuban@iitm.ac.in}
\affiliation{Department of Applied Mechanics, IIT Madras, Chennai 600036, India}

\author{R. Rajesh}
\email{rrajesh@imsc.res.in}
\affiliation{Institute of Mathematical Sciences, C.I.T. Campus, Taramani, Chennai 600113, India
}
 \altaffiliation[Also at ]{Homi Bhabha National Institute, Training School Complex, Anushakti Nagar, Mumbai 400094, India}%Lines break automatically or can be forced with \\

 %\altaffiliation[Also at ]{Department of Applied Mechanics, IIT Madras, Chennai 600036, India}%Lines break automatically or can be forced with \\

\date{\today}% It is always \today, today,

\begin{abstract}
Fracture processes in multi-phase solids are inherently complex due to multiple competing mechanisms. Here, we investigate the elastic and fracture behaviour of two-phase solids—comprising a fragile phase and a tough phase using a disordered spring network model. The macroscopic response is found to depend on the failure strain mismatch, the elastic modulus ratio, as well as the relative composition of the constituent phases. As the proportion of the tough phase increases, the system undergoes a reentrant phase transition in fracture behaviour: from brittle to ductile-like and back to brittle. These transitions are identified through both avalanche statistics and cluster size characteristics of broken springs. Notably, the avalanche exponent associated with the majority phase changes universality class during the brittle-to-ductile transition. Analysis of time evolution of cluster characteristics reveals distinct growth mechanisms in the two regimes. In the brittle regime, dominant clusters rapidly absorb other large clusters, keeping the total number of clusters nearly constant. In contrast, the ductile regime is characterised by more gradual coalescence, leading to a decrease in the total number of clusters over time while their average size increases. We provide a physical interpretation of the mechanisms underlying the observed switch in fracture behaviour.
\end{abstract}

\maketitle

\section{\label{sec: Int}Introduction}
%%%% Insert A head here
Structural materials—such as metal alloys, concrete, and composite materials—as well as biological materials like bone, wood, and nacre, are inherently multi-phase in nature~\cite{meyers2008biological,jiang2025effect,niu2022superior,qiu2024scalable,gerhard2017design}. Fracture processes in such materials are complex, involving several competing mechanisms. The dominance of any particular mechanism depends not only on the fracture properties of the individual phases and their interfacial characteristics but also on their relative proportions~\cite{koester2008effect,sarikaya1989mechanical,koester2008true,bechtle2010crack,smith1983lower,jones2018mechanics,nikolic2018lattice}. Depending on the scale of the system under consideration, particularly in relation to the characteristic length scale of heterogeneity, different modelling approaches are appropriate. For systems much larger than the heterogeneity scale, the effect of heterogeneity can be incorporated by using an effective single-phase constitutive response with embedded disorder~\cite{kun2024failure,mayya2016splitting,mayya2017role,perdahciouglu2011constitutive}. However, when the system size is comparable to the heterogeneity scale, it becomes essential to model the phases explicitly.~\cite{yongqiang2003theoretical,urabe2010fracture,senapati2023role,parihar2020role,brely2015hierarchical,yip2006irregular,noguchi2024fracture,tauber2020microscopic}. This enables a more realistic investigation of the fracture process and its influence on macroscopic material properties such as strength, toughness, and ductility. Findings from predictive simulations of multi-phase materials are vital not only for life assessment of existing structural composites but also for understanding the exceptional performance of natural bio-composites. These insights potentially can, in turn, inform the design of advanced, bio-inspired materials.

In fracture experiments, the time evolution of damage, whether through nucleations of micro-cracks, stable growth of micro-cracks, or rapid coalescence, releases intermittent elastic stress waves. These stress waves can be captured as acoustic emission (AE) signals, allowing damage to be quantified in terms of various AE parameters, such as amplitude, energy, duration, and count of the digitised AE waveform~\cite{senapati2024acoustic,chakraborty2024acoustic,saha2022statistical,chen2020avalanches,nataf2014avalanches}. The amplitude ($A$) of these AE events exhibits a power-law distribution of the form $P(A) \sim A^{-\tau_A}$, where $P(A)$ is the probability of an amplitude ($A$) and the exponent $\tau_A$ varies depending on the material. For example, the exponent is 1.95 for cellular glass~\cite{maes1998criticality}, 2.0 for volcanic rocks~\cite{diodati1991acoustic}, 0.52-0.84 for concrete~\cite{petri1994experimental}, 1.3-1.4 for porcine bone~\cite{baro2016avalanche}, 2.0 for PUFOAM~\cite{deschanel2009experimental}, 1.83 for human teeth~\cite{wang2021cracking}, 2.0 for glassy epoxy~\cite{senapati2024acoustic}, 1.6-2.0 for paper~\cite{rosti2010statistics}, and 2.0 for rock~\cite{jiang2017predicting} . The power law behaviour suggests the absence of a characteristic event size, consistent with critical-like dynamics observed in disordered systems. However, variations in the avalanche exponent across different materials indicate that it is not universal for all disordered solids.

Statistical fracture models, such as the Fiber Bundle Model (FBM), Fuse Network Model, and Random Spring Network Model (RSNM) have played a crucial role in capturing several key features of the fracture process in disordered solids~\cite{alava2006statistical,zapperi1997plasticity,hansen1994burst,kun2000damage,hidalgo2002fracture,kumar2022interplay}. These models effectively reproduce the nucleation of independent micro-cracks, their subsequent growth and interactions, the intermittent nature of damage evolution, the emergence of complex fracture paths, and statistical signatures such as power-law distributions~\cite{pradhan2005crossover,hidalgo2002fracture,kun2008universality,andersen1997tricritical,moreno2000fracture,sinha2020phase}. More recently, statistics of super events of the time series have been analysed to predict imminent final failure~\cite{senapati2024record} as well as to characterise disorder in a glassy epoxy-based polymer~\cite{senapati2024acoustic}. These studies treated the heterogeneous solid as a single phase with disorder that incorporates distributed fracture characteristics.

Comparatively fewer studies have investigated in as much detail the fracture behaviour of multi-phase solids where the phases are explicitly modelled. In one of the earliest studies of mixed phases using FBM, a discontinuous uniform distribution for failure threshold was used. Existence of non-universal, non-mean field behaviour was shown for certain fraction of phases~\cite{divakaran2007effect}. Further using FBM, an idealized two-phase solid is modelled with one phase being unbreakable~\cite{hidalgo2008universality}. The analytical solutions show that the introduction of the unbreakable phase results in a switch in the mean field exponent from its well established value for a disordered single phase solution, $5/2$, to $9/4$. This switch occurs when the fraction of second phase reaches a critical limit $0.5$. Further increase in the fraction of the second phase results in a distribution with the same new exponent but with a decreasing cut-off. The switch is also shown to be conditioned to disorder distributions for which the constitutive curve has a peak and an inflection point. Applying local load sharing rules within FBM, the critical limit for the fraction of the second phase is seen to shift to $\approx 0.06$.  Among stochastic fracture models, RSNM takes the mechanics based approach to find the load distribution and state of deformation and damage, particularly suited for simulating more realistic macroscopic responses~\cite{curtin1990brittle,urabe2010fracture}, including pre- and post-peak behaviour~\cite{bolander2005irregular,yip2006irregular,wang2020lattice}, complex crack paths~\cite{zapperi1997plasticity,liu2007lattice,brely2015hierarchical,tauber2020microscopic}, the power-law scaling of avalanche distributions~\cite{shekhawat2013damage,kumar2022interplay,mayya2016splitting,ray2006breakdown}, as well as the dynamics of record events that characterise the acceleration in damage growth~\cite{senapati2024acoustic,senapati2024record}. Using RSNM for a two-phase domain, Noguchi et al. ~\cite{noguchi2024fracture} modelled a mix of breakable and unbreakable phases, where the unbreakable phase is architectured as a frame of square-like cells. Authors argue that, for the three configurations of the frame considered, the burst size distribution exhibits power law with a universal exponent, close to that for single phase RSNM, and only the exponential cutoff is shown to be dependent on the internal architecture. Studies on statistics of fracture events in a mixed phase have not explored the possibility of both phases being breakable and mismatched in both their elastic and fracture characteristics. 

In a solid of two phases, mismatched in their elastic as well as fracture characteristics, if the individual phases exhibit power-law with a universal constant, is there a switch in the power law exponent to another universality class? If so, then is the critical fraction of one phase in the other phase same when the phases are interchanged and whether the cause of the double switches has similar mechanisms? In the present work, using a RSNM, we simulate the damage and fracture behaviour of a two-phase solid. To model the two-phase solid, the elastic behaviour of the springs is chosen to be either hard or soft, similarly, the fracture behaviour is chosen to be either fragile or tough [see Fig.\ref{figM2}]. We find that the shift in fracture behaviour from brittle (abrupt load drop post-peak) to ductile-like (gradual load drop post-peak) and back to brittle is dependent on both the elastic heterogeneity and the relative proportion of the phases. We develop a discussion on the unique damage mechanisms responsible for this re-entrant transition in fracture behaviour. Despite the distinct damage mechanisms governing the transition, the avalanche distribution of the phases consistently shifts from one universality class to another. The spatial interaction of the damage activity leading to final failure is also explained using the dynamics of the damage cluster of the system.

\section{\label{sec:Mod} Model}
\subsection{\label{sec: Mod1} The Random Spring Network Model}

\begin{figure}
\centering
\includegraphics[width=9.5cm, height=5.34cm]{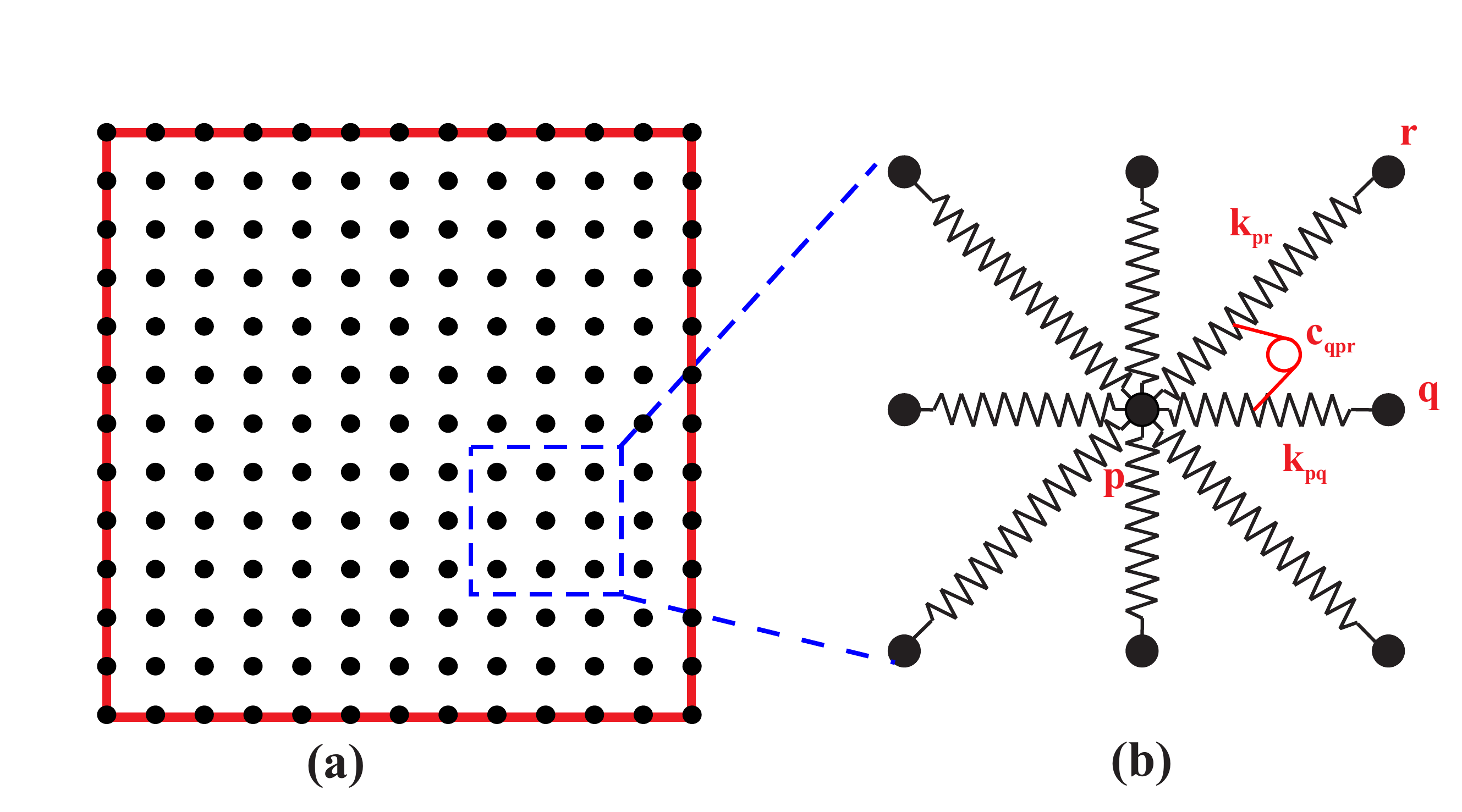}
\caption{ (a) Discretisation of a continuum domain. (b) The spring network.}
\label{figM1}
\end{figure}

A square domain of a two-phase solid is discretised by a $100\times100$ square lattice, with a lattice spacing $a_0 = 0.5 mm$, as shown in Fig.~\ref{figM1}(a). The lattice point $(p)$ in a unit lattice is connected with its nearest $(q)$ and next-nearest $(r)$ neighbours through extensional springs of stiffness  $k_{pq}$ and $k_{pr}$, respectively. It interacts rotationally with its two adjacent neighbours $(q)$ and $(r)$ through rotational spring of stiffness $c_{qpr}$, as shown in Fig.~\ref{figM1}(b).

When the network deforms, the potential energy $\phi$ is stored in both extensional and rotational springs and has components $\phi_{ext}$ and $\phi_{rot}$,

\begin{equation}
\label{eq1}
\phi = \phi_{ext}+\phi_{rot}.
\end{equation}
The extensional component of the potential energy is:
\begin{equation}
\label{eq2}
\phi_{ext} = \sum_{\left\langle ij\right\rangle}\frac{1}{2}k_{ij}\left(|\vec{r}_{i}-\vec{r}_{j}|-a_{ij}\right)^2,
\end{equation} 
where the sum is over all extensional springs, $a_{ij}$ is the undeformed distance between lattice points $i$ and $j$, $\vec{r}_{i}$ and $\vec{r}_{j}$ are the position vectors of their deformed configuration and $k_{ij}$ is the stiffness of the extensional spring connecting them. The rotational component of the potential energy in the lattice is:
\begin{equation}
\label{eq3}
\phi_{rot} = \sum_{\left\langle pqr\right\rangle}\frac{1}{2}c_{pqr}\left(\theta_{pqr}-\frac{\pi}{4}\right)^2,  
\end{equation}
where the sum is over all 8 triplets of lattice points connected by torsional springs, $\theta_{pqr}$ denotes the angle between two adjacent neighbours $q$ and $r$ with the lattice point $p$ in the deformed configuration, and $c_{pqr}$ denotes the stiffness of the rotational spring. In the undeformed lattice configuration $\theta_{pqr} = \pi/4$. The net potential energy density of the lattice, as defined in Eq.(\ref{eq1}), is equated with the strain energy density of the elastic continuum to express the elastic continuum parameters, Elastic modulus $E$ and Poisson's ratio $\nu$ in terms of lattice parameters as~\cite{monette1994elastic}:
\begin{equation}
\label{eq4} 
E = \frac{8k\left(k+\frac{c}{a_0^2}\right)}{3k+\frac{c}{a_0^2}},
\end{equation}

\begin{equation}
\label{eq5}
\nu = \frac{\left(k-\frac{c}{a_0^2}\right)}{3k+\frac{c}{a_0^2}},
\end{equation} 
where, the stiffness of horizontal and  vertical springs is $2k$, that of diagonal spring is $k$, and $c_{pqr}=c$.

The system is pulled in the vertical direction at a strain rate of $10^{-5}$ per loading step. To avoid free-body displacement, the bottom left lattice point is kept fixed and the other lattice points in the bottom row are allowed to move only in the horizontal direction. With every strain increment, the springs in the network deform, and the system evolves iteratively to attain the state of quasi-static equilibrium. In each iterative step, the updated position vector $\vec{r}_p(t + \Delta t)$ of the lattice points are computed based on the position vector of the last two time steps $\vec{r}_p(t)$ and $\vec{r}_p(t - \Delta t)$ using Verlet algorithm~\cite{verlet1967computer} as:
\begin{equation}
\label{eq7} 
\vec{r}_p(t \!+ \!\Delta t) = \vec{r}_p(t)\left(2\!-\!\gamma\Delta t\right)-\vec{r}_p(t - \Delta t)\left(1\!-\!\gamma\Delta t\right)+ \vec{a}_p (\Delta t)^2.
\end{equation} 
 At each lattice point, the resultant force components of the reaction forces are calculated using
\begin{equation}
\label{eq6} 
\vec{a}_{p}= -\nabla_{\vec{r}_p}\phi -\gamma\vec{v}_p,
\end{equation}
where mass is set to unity. To achieve quasi-static equilibrium, an additional dissipative term $-\gamma\vec{v}_p$ is introduced, with $\gamma$ set to 0.8. The velocity in the dissipative term is calculated using the backward difference formula $\vec{v}_p(t)=[r(t)-r(t-\Delta t)]/\Delta t +O(\Delta t)$. The system is assumed to reach quasi-static equilibrium when the kinetic energy of each particle is below a predefined threshold. This is re-verified by checking the forces in the top and bottom edges are equal within a pre-defined tolerance of $0.99< \lvert \frac{F_{top}}{F_{bot}} \rvert < 1.01$. At the end of equilibration, if any of the surviving springs fail, based on the failure criteria discussed below, then the system is again re-equilibrated following the same iterative process until no further breakage of springs occurs. We note that when an extensional spring fails, as a consequence, its associated torsional springs are also taken to have failed.

\subsection{\label{sec: Mod2} Modelling of two-phase solids}

\begin{figure}
\centering
\includegraphics[width=9cm, height=9cm]{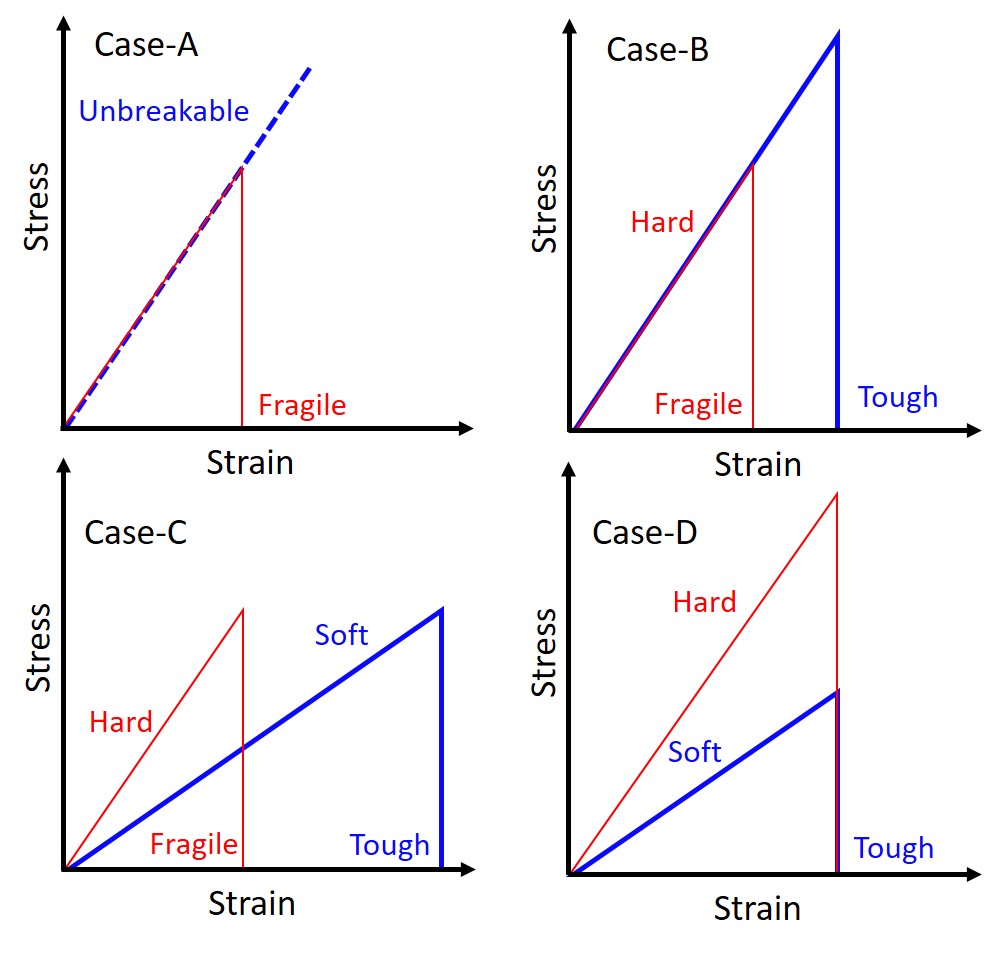}
\caption{Schematic representation of constitutive and failure behaviour of the constituent phases in two-phase solids with Case-(A) $\alpha = 1$, $\beta = \infty$, Case-(B) $\alpha = 1$, $\beta = 20$, Case-(C) $\alpha = 20$, $\beta = 20$ and Case-(D) $\alpha = 20$, $\beta = 1$. Note that when $\alpha = \beta = 20$, the failure stress threshold is the same for both phases.}
\label{figM2}
\end{figure}

In this study, each phase of the two-phase solid is taken to be distinct in terms of its elastic modulus and the failure strain threshold. The fracture process in the two-phase system, thus, has a coupled effect of elastic mismatch, different failure threshold strains, as well as the relative proportion of each phase. The phase with the lower failure strain is defined as the fragile phase, while the phase with the higher failure strain is defined as the tough phase. To quantify the elastic mismatch between these phases, we introduce the elastic modulus ratio,
\begin{equation}
\label{eqconstit} 
\alpha = \frac{E_{fragile}}{E_{tough}},
\end{equation}
 where $E_{tough}$ and $E_{fragile}$ are the elastic moduli of the tough and fragile phases, respectively. Similarly, the mismatch between the mean failure strains of the constituent phases is represented by the ratio,
 \begin{equation}
\label{eqcFail} 
 \beta = \frac{\langle\epsilon_f \rangle _{tough}}{\langle \epsilon_f \rangle _{fragile}},
\end{equation}
 where $\langle \epsilon_f \rangle _{tough}$ and $\langle \epsilon_f \rangle _{fragile}$ are the mean failure strains of the tough and fragile phases, respectively.

In the first part of the investigation, to focus on the impact of difference in failure strain thresholds between the constituent phases, they are taken to be elastically identical ($\alpha = 1$). We consider case $A$ where the tough phase is taken to be unbreakable [Fig.~\ref{figM2}(a)] and case $B$ where the tough phase is breakable with $\beta = 20$ [Fig.~\ref{figM2}(b)]. For these analyses, the elastic modulus for both phases is set to $E_{tough} = E_{fragile} = 100$ GPa. In the second part of the investigation, we examine the combined influence of differences in failure strain thresholds ($\beta = 20$) as well as the mismatch in elastic moduli of the two phases ($\alpha = 20$), as shown in Fig.~\ref{figM2}(c). Finally, we also consider the cases where the strain thresholds of the two phases are equal ($\beta = 1$), but the phases are elastically mismatched ($\alpha = 20$) [see Fig.~\ref{figM2}(d)]. In all cases, the Poisson's ratio is taken to be $\nu=0.3$ for both phases.

Disorder in the model is incorporated by taking the spring failure strain thresholds of the corresponding phases to be normally distributed with their mean values $\langle \epsilon_f 
\rangle _{fragile} = 0.0125$ and $ \langle \epsilon_f \rangle _{tough} = 0.25$ and a common standard deviation of $25\%$ of the respective mean for all phases. In the simulations, a spring is taken to fail if its strain exceeds its assigned threshold value.

The spring constants for each phase are first calculated using Eqs.~(\ref{eq4}) and (\ref{eq5}) by assuming a homogeneous system composed of only that phase. Then, the spring constants of the four springs connecting a lattice site to its south, southwest, west, and northwest neighbours are assigned based on the phase of that site. Similarly, the spring constant for the eight rotational springs interacting with the site is also determined and assigned. This method ensures a consistent and unique assignment of spring constants for each lattice site in the two-phase solid. To investigate the effect of heterogeneity on the fracture response, we have chosen $r \in [0.0, 1.0]$, where $r$ represents the relative proportion of the tough phase in both elastically matched and mismatched solids. When the strain thresholds are equal (Fig.~\ref{figM2}(d)) then $r$ refers to the fraction of soft phase.

\section{\label{sec:Result} Results}

\subsection{\label{sec:1} Effect of unbreakable tough phase in elastically matched solids}
This section explores the effect of a significantly tougher phase on the fracture behaviour of the fragile phase and its manifestation in the response of a heterogeneous solid. To isolate this effect, the tough phase is assumed to be unbreakable. We first discuss the effect of the relative proportion of the unbreakable tough phase on the overall constitutive behaviour, followed by its effect on the avalanche distribution and conclude with a discussion on the spatial evolution of damage.

\subsubsection{\label{sec:1.1}The constitutive behaviour}

\begin{figure}[!h]
\centering
\includegraphics[width=7.0cm, height=7.0cm]{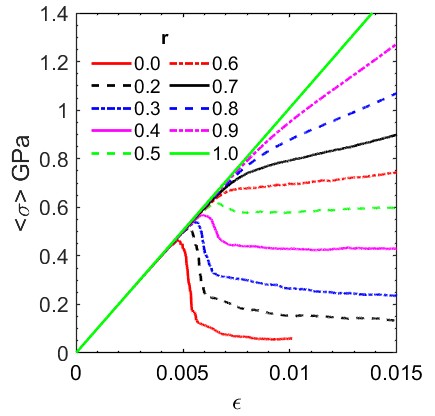}
\caption{Effect of fraction of unbreakable phase on the constitutive response of elastically matched solids ($\alpha = 1$, $\beta = \infty$).}
\label{fig1}
\end{figure}

The constitutive behaviour of two-phase systems for a range of proportions, starting from a breakable homogeneous fragile phase ($r = 0$) to a fully unbreakable tough phase ($r = 1.0$), having the same elastic modulus of $E = 100$ GPa for both phases, is shown in Fig.~\ref{fig1}. In the homogeneous fragile phase ($r = 0$), the slope of the stress-strain curve remains constant, almost up to its strength, followed by a sharp load drop like in brittle fracture. In contrast, in the tough phase ($r = 1.0$), the slope is maintained throughout without any load drop, reflecting the unbreakable nature of springs. In the range $r$ = (0.0,0.5], the constitutive behaviour is initially linearly elastic, having the same $E$ as the homogenous phases, followed by a softening behaviour post-peak when the load drops rapidly. Further deformation asymptotically tends to elastic deformation of the remaining unbreakable springs. For higher $r$, the extent of softening reduces and nearly vanishes for $r = 0.5$. In this range of $r < 0.5$, the strength,  $\sigma_c$, and its corresponding strain $\epsilon_{\sigma_c}$ increase monotonically with $r$. For $r > 0.5$, the constitutive behaviour remains unchanged almost up to the linear limit, followed by a hardening response. For higher $r$, the stress at the linear limit is higher. After most fragile springs fail, the network resumes linear elastic behaviour but with reduced effective elastic modulus. For $r = 1.0$, since there are no breakable springs, there is no deviation from the initial elastic behaviour. 

The above observation may be rationalised as follows. In a heterogeneous system, as damage propagates within the fragile phase, the compliance of the network increases. With continued loading, further failures in the fragile springs progressively diminish its load-carrying capacity. Consequently, the unbreakable tough phase carries progressively increasing share of the applied load. As the proportion of the tough phase, $r$, increases, it bears most of the applied load from the beginning, resulting in either a gradual or no significant load drop. Consequently, after an initial common linear elastic response across all heterogeneous cases, large load drops that are characteristic of brittle fracture are observed only up to a threshold, $r_c = 0.4$. 

\subsubsection{\label{sec:1.2}Avalanche distribution}

\begin{figure}
\centering
\begin{subfigure}[b]{0.45\textwidth}
\centering
\includegraphics[width=6.5cm, height=6.5cm]{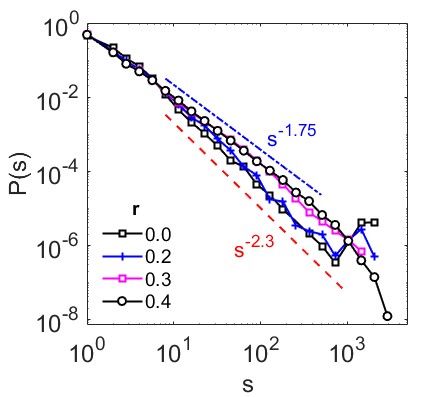}
\caption{}
\label{fig2a}
\end{subfigure}
\hfill
\begin{subfigure}[b]{0.45\textwidth}
\centering
\includegraphics[width=6.5cm, height=6.5cm]{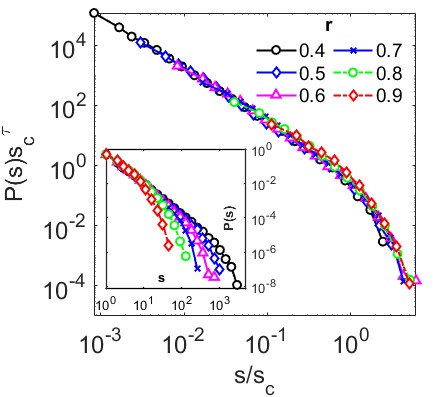}
\caption{}
\label{fig2b}
\end{subfigure}
\begin{subfigure}[b]{0.45\textwidth}
\centering
\includegraphics[width=6.5cm, height=6.5cm]{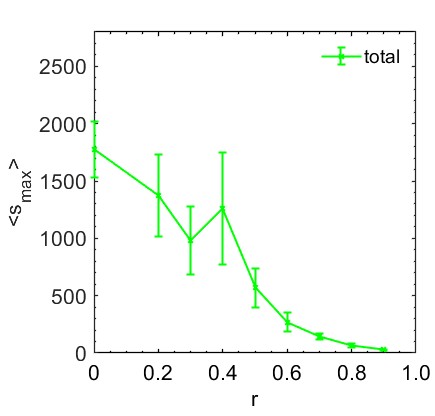}
\caption{}
\label{fig2c}
\end{subfigure}
\hfill
\begin{subfigure}[b]{0.45\textwidth}
\centering
\includegraphics[width=6.5cm, height=6.5cm]{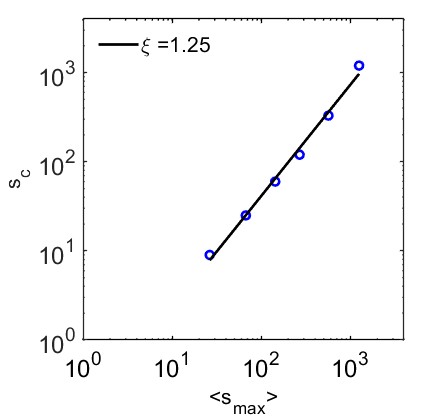}
\caption{}
\label{fig2d}
\end{subfigure}
\caption{Avalanche distribution in elastically match solids ($\alpha = 1$, $\beta = \infty$) with unbreakable tough phase for (a) $r \leq 0.4$ and (b) $r \geq 0.4$. In (b), the rescaled avalanche distributions are shown in the main plot, with unscaled distributions in the inset. (c) The variation of the average maximum avalanche size $\langle s_{max} \rangle $ with $r$, and (d) The scaling of $\langle s_{max} \rangle$ with the cutoff $s_c$ obtained from the data using Eq.(\ref{eq8}) for $r \geq 0.4$.}
\label{fig2}
\end{figure}

To characterise the fracture process, we next examine the avalanche statistics associated with the breakage of springs. An avalanche of size $s$ is typically defined as the number of springs that fail within a given increment of applied strain. The probability distribution, $P(s)$, of avalanche sizes $s$ for a moderately disordered single phase RSNM is known to follow a power-law form, $P(s) \sim s^{-\tau}$, with exponent, $\tau \approx 2.3$~\cite{senapati2024acoustic,kumar2022interplay,ray2006breakdown,zapperi1997first}. The disorder in the system was introduced through the distributed nature of the failure threshold assigned to individual springs. 

In a two-phase solid, to examine the effect of the relative proportion of the phase on the exponent of the avalanche distribution, we examine the distributions for $r \in [0, 0.4]$ and  $r \in [0.4, 0.9]$ in Figs.~\ref{fig2a} and ~\ref{fig2b}, respectively. In Fig.~\ref{fig2a} for $r = 0.0$, the avalanche exponent $\tau$ of the homogeneous fragile phase is 2.3, which is consistent with the available literature for RSNM. Around $r = 0.3$, the exponent switches to $1.75$.  This switch can be understood as follows. In two-phase solids, presence of the unbreakable phase restricts the size of the largest avalanche prior to failure, which would typically results in triggering the final failure. This restriction delays and, in some cases, prevents the onset of catastrophic fracture. Consequently, larger avalanches continue to form and their size is only constrained by the cluster-size distribution of the breakable phase. This behaviour is reflected in Fig.~\ref{fig2a}, where the probability of larger avalanches is higher in two-phase systems compared to the single phase (excluding the avalanche at failure), resulting in a smaller power law exponent. 

When the ratio $r$ exceeds the critical limit $r_c$, the bump in $P(s)$ associated with large catastrophic avalanches is replaced by an exponential cutoff, as shown in Fig.~\ref{fig2b}. Further increase in $r$ does not alter the exponent of the power-law distribution, thus, establishing two distinct universality classes, though the characteristic avalanche size $s_c$ decreases (see inset of Fig.~\ref{fig2b}). Scaling analysis reveals that by rescaling both axes with an appropriate exponent and the characteristic avalanche sizes, these avalanche distributions can be collapsed onto a master curve of the form:
\begin{equation}
\label{eq8}
{s_c}^{\tau}P(s) \sim {(s/s_c)}^{-\tau}\exp(-s/{s_c}),
\end{equation}
 with $\tau \approx 1.75$. 

Next, we look at the largest avalanche in the simulation which is indicative of the extent of rapid growth of damage. In Fig.~\ref{fig2c}, for $r = 0.0$, $\langle s_{max}\rangle$ is maximum, and with increasing fraction of unbreakable springs, the largest avalanche size reduces. The decrease with increasing $r$ suggests that much larger defects are present prior to rapid growth of damage in form of the largest avalanche. Thus, the sudden growth is more localised with a comparatively smaller single largest avalanche. Beyond $r = 0.4$, $\langle s_{max}\rangle$ decreases suddenly because the domain has isolated regions of breakable springs, which automatically results in limiting the maximum avalanche size. Additionally, a scale-free relationship exists between $s_c$ and the average maximum avalanche size $\langle s_{max}\rangle$, given by $s_c \sim \langle s_{max} \rangle ^\xi$ with $\xi = 1.25$ as shown in Fig.~\ref{fig2d}. This value of $\xi$, interestingly, coincides with that found in fiber bundle model~\cite{kovacs2013brittle}.

\subsubsection{\label{sec:1.3}Microstructure of damage}

\begin{figure}
\centering
\begin{subfigure}[b]{0.95\textwidth}
\centering
\includegraphics[width=10.0cm]{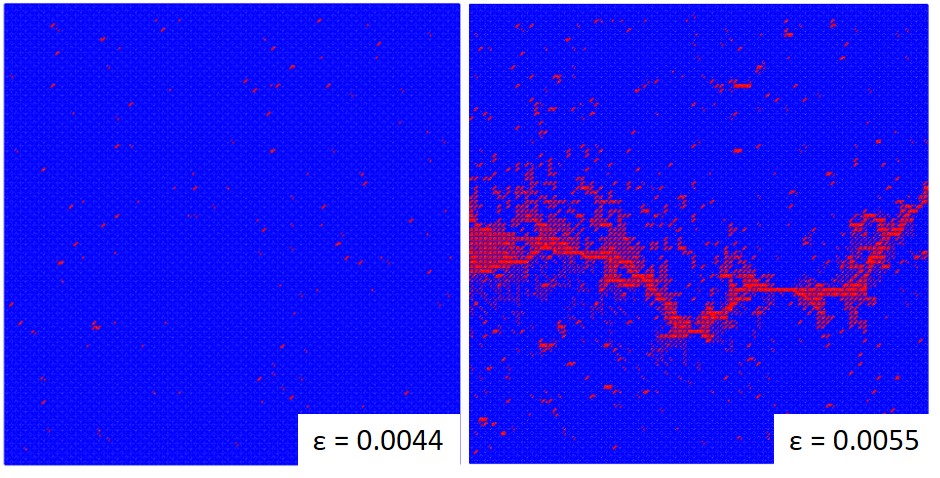}
\caption{}
\label{fig3a}
\end{subfigure}
\begin{subfigure}[b]{0.95\textwidth}
\centering
\includegraphics[width=10.0cm, height=10.0cm]{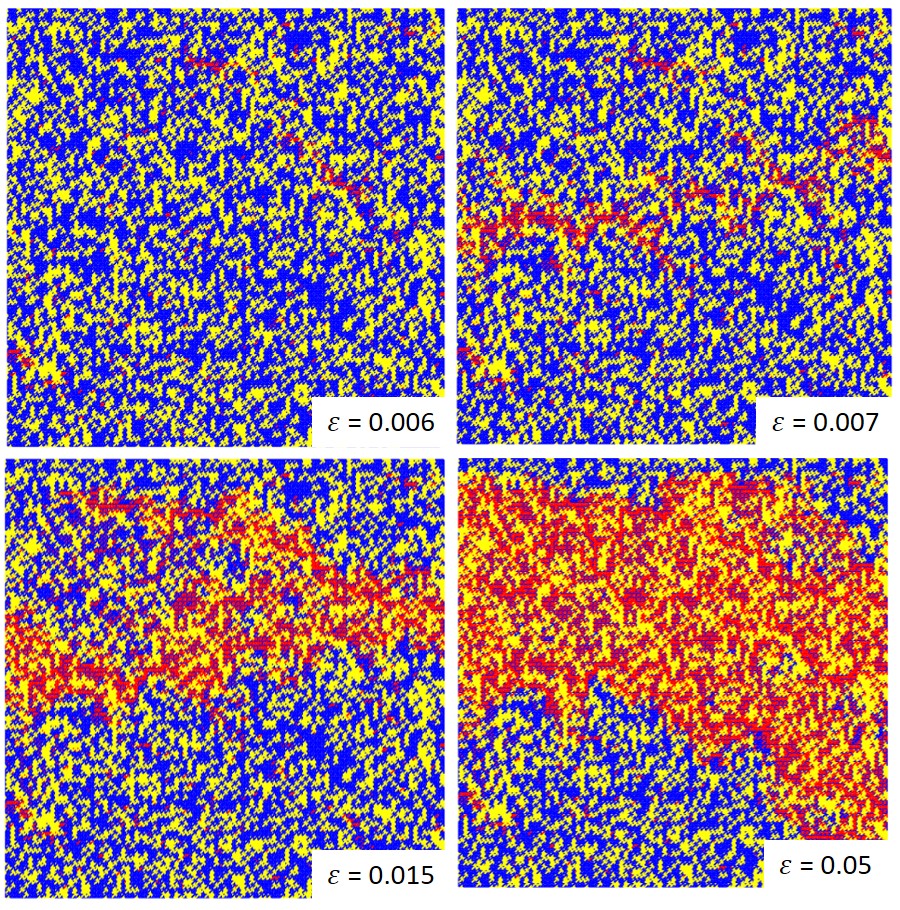}
\caption{}
\label{fig3b}
\end{subfigure}
\caption{Damage evolution during various stages of applied strain for relative proportion (a) $r = 0$ and (b) $r = 0.4$. Blue (grey), yellow (light grey), and red (dark grey) colours represent intact breakable, unbreakable, and broken breakable springs, respectively. The snapshots are for $\alpha=1$, $\beta=\infty$.}
\label{fig3}
\end{figure}

In a homogeneous breakable system with $r = 0$, initially micro-cracks nucleate randomly throughout the domain and grow independently, without spatial correlation. These micro-cracks typically form small clusters (connected failed springs) that are much smaller than the system size (see Fig.~\ref{fig3a}, $\epsilon = 0.0044$). As the system is further loaded, a spanning macro crack rapidly forms, leading to catastrophic failure, as can be seen in Fig.~\ref{fig3a}, ($\epsilon = 0.0055$). In contrast, in a two-phase system with $r > r_c$, the growth and interaction of these clusters exhibit distinct behaviour, as illustrated in Fig.~\ref{fig3b}. In these systems, micro-cracks begin to form at an early stage of loading ($\epsilon = 0.006$) as for a single phase. As loading continues, however, these micro-cracks grow and merge into larger clusters, which are larger than those observed in the brittle fracture of the fragile phase. By ($\epsilon = 0.05$), a dominant spanning cluster emerges,  covering nearly the entire domain by connecting almost all the broken springs.

\begin{figure}
\centering
\includegraphics[width=7.0cm, height=7.0cm]{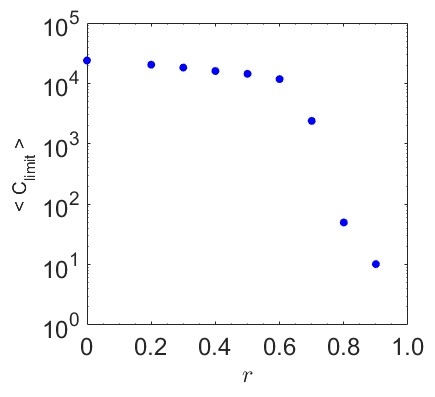}
\caption{The variation of the mean cluster size, $\langle C_{limit} \rangle$, of the breakable springs in the initial configurations, with $r$. The data are for $\alpha = 1$, $\beta = \infty$.}
\label{fig4}
\end{figure}

The mechanisms of fracture are also manifested in the time evolution of clusters. We now examine the manifestation of the transition in the fracture mechanism on the strain-dependent cluster size distribution of the broken springs. The average cluster size $\langle C_{av} \rangle$ is defined as the ratio of the second moment to the first moment of the cluster sizes $C_{i}$ of the broken springs:
\begin{equation}
\label{eq9}
\langle C_{av} \rangle = \frac{\sum_i {C_i}^2}{\sum_i {C_i}},
\end{equation}
where the average is computed over several realisations. Prior to examining the time evolution, however, to understand the configurational effect of the unbreakable phase, we first examine the effect of relative proportion of the tough phase $r$, on the largest possible cluster size $\langle C_{limit} \rangle$, obtained from Eq.(\ref{eq9}) by considering all fragile springs to be broken. Clearly, $\langle C_{limit} \rangle$ represents an upper bound for $\langle C_{av} \rangle$. The variation of $\langle C_{limit} \rangle$ with $r$ is shown in Fig.~\ref{fig4}. 
$\langle C_{limit} \rangle$ decreases sharply  after $r^*\approx 0.6$. We note that the bond percolation threshold for a square lattice with nearest and next-nearest neighbours is $\approx 0.25$ ~\cite{feng2008percolation,ouyang2018equivalent,xu2021critical}, corresponding to $r^* \approx 0.75$.  However, tough and fragile bonds are assigned to the RSNM in a correlated fashion, resulting in an $r^*$ different from $0.75$. For $r < r^*$, there is a spanning cluster of fragile springs, leading to system size dependent $\langle C_{limit} \rangle$.

\begin{figure}
\centering
\includegraphics[width=10.0cm, height=9.0cm]{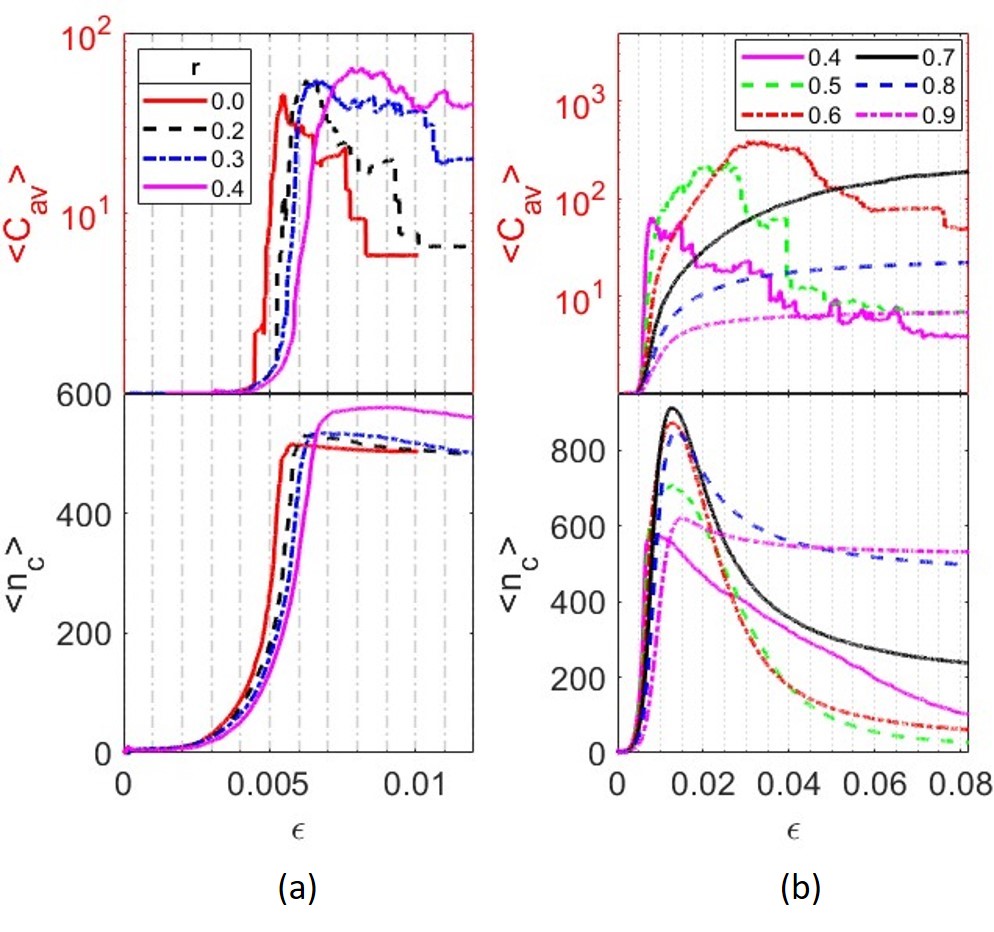}
\caption{The variation of average cluster size $\langle C_{av} \rangle $ of the broken springs and average number of clusters $\langle n_c \rangle $ with $\epsilon$ for various proportions described by $r$. Note that the largest cluster is omitted in the calculation of $\langle C_{av} \rangle $ for this figure.  The data are for $\alpha=1$, $\beta=\infty$.}
\label{fig5}
\end{figure}

Understanding the constraint of the unbreakable phase on the size of the clusters, we explore the dynamics of $C_{av}$ for various proportions $r$, which will help to characterise the growth rate of small cluster (micro-cracks), the initiation of the dominant spanning cluster, and its subsequent growth. In the fracture process, the dominant spanning cluster, once formed, is so large that the merging of smaller clusters has no noticeable effect on its size. To retain the details of the growth of the cluster population, we exclude the largest cluster in the calculation of $C_{av}$~\cite{kovacs2013brittle}. 

In Fig.~\ref{fig5}, there are three regimes of relative proportion that have distinct time evolution of cluster characteristics. In the breakable phase dominant regime, for $r < r_c$, shown in~\ref{fig5}(a), with increase in applied strain the average cluster size and the number of clusters, $\langle n_c \rangle $, initially increase at a slow rate consistent with independent nucleation of clusters. As applied strain approaches a critical limit, both $\langle C_{av} \rangle $ and $\langle n_c \rangle $ show a sudden rise, suggesting rapid nucleation of new clusters as well as the growth of the existing ones. Both $\langle C_{av} \rangle $ and $\langle n_c \rangle $ peak at the same strain, which also corresponds to peak in the macroscopic response. Further increase in applied strain, shows rapid decrease in $\langle C_{av} \rangle $ and negligible decrease in $\langle n_c \rangle $, implying that the larger clusters get absorbed into the single dominant cluster.

In the mixed regime, $r_c<r<r^*$, as shown in~\ref{fig5}(b),  with increasing strain $\langle n_c \rangle $ peaks before $\langle C_{av} \rangle $, suggesting that the nucleation activity subsides and coalescence begins. The average cluster size, however, continues to increase as clusters continue to merge, and peak for $\langle C_{av} \rangle $, an order of magnitude larger, is reached later. Occurrence of peak in $\langle C_{av} \rangle $ is again a signature of formation of a dominant cluster which absorbs other large clusters, however, in this regime the average cluster decreases comparatively slower.

In the unbreakable phase dominant regime, $r>r^*$, also shown in~\ref{fig5}(b), after the initial independent nucleation and growth of clusters, $\langle n_c \rangle $ reaches a peak suggesting onset of coalescence. $\langle C_{av} \rangle $, while increases, never attains a peak as the over riding presence of unbreakable springs disallows formation of a single dominant cluster.

\subsection{\label{sec:2} Effect of discontinuous threshold strain distribution}
We now examine the additional features in the fracture behaviour of a two-phase solid when the tough phase, previously unbreakable, is now breakable. Initially, we consider the simpler case of elastically matched system where both fragile and tough phases have the same elastic properties ($\alpha = 1$) but different mean failure strain thresholds ($\beta = \frac{\langle \epsilon_f \rangle _{tough}}{\langle \epsilon_f \rangle _{fragile}}$= 20), as shown in Fig.~\ref{figM2}(b).

\subsubsection{\label{sec:2.1}The constitutive behaviour}

\begin{figure}
\centering
\begin{subfigure}[b]{0.45\textwidth}
\centering
\includegraphics[width=6.5cm, height=6.5cm]{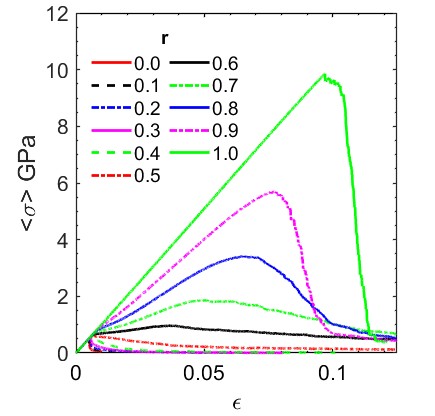}
\caption{}
\label{fig6a}
\end{subfigure}
\hfill
\begin{subfigure}[b]{0.45\textwidth}
\centering
\includegraphics[width=6.5cm, height=6.5cm]{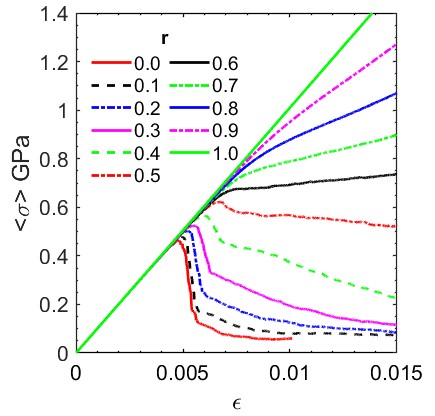}
\caption{}
\label{fig6b}
\end{subfigure}
\begin{subfigure}[b]{0.45\textwidth}
\centering
\includegraphics[width=6.5cm, height=6.5cm]{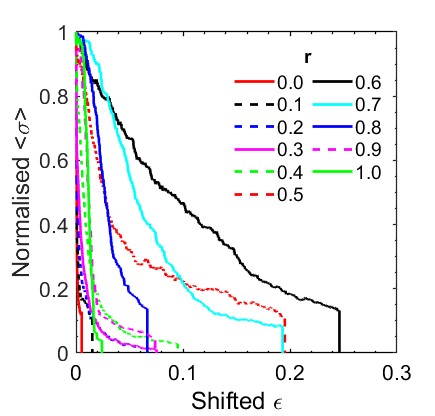}
\caption{}
\label{fig6c}
\end{subfigure}
\caption{Constitutive behaviour of elastically matched solids ($\alpha = 1$, $\beta = 20$) for various proportions, $r$, of the breakable tough phase. The complete behaviour up to the final fracture is in (a), and the zoomed view for the small deformation regime is in (b). To clearly illustrate the rate of load drop, the normalised stress-strain behaviour post-peak is shown in (c). For each $r$, $\langle \sigma \rangle $ is normalised by its corresponding peak value, and $\epsilon$ is shifted by subtracting the strain value at maximum $\langle \sigma  \rangle$.}
\label{fig6}
\end{figure}

The stress-strain response, shown in Fig.~\ref{fig6a}, when compared to that of the two-phase system with an unbreakable tough phase, Fig.~\ref{fig1}, differs only in the later stage of loading, as damage in the tough phase occurs only at larger strains. In fragile phase-dominated solids ($r \leq 0.2$), the fragile phase primarily carries the applied load until final failure. Consequently, whether the tough phase is breakable or not has a negligible impact on the macroscopic response. As the proportion of the breakable tough phase increases, the behaviour remains unchanged until the peak, as shown in Fig.~\ref{fig6b}. In the post-peak stage, the macroscopic response is primarily governed by the damage characteristics of the breakable tough phase. 

The normalised stress-strain behaviour beyond the strength is shown in Fig.~\ref{fig6c}. It reveals that the load drops sharply in two-phase solids with a smaller proportion of the tough phase ($r \leq 0.3$), showing brittle fracture behaviour. On further increase in the proportion of tough phase, the solid exhibits gradual load drop, showing ductile fracture behaviour in the range ($r \in [0.4, 0.8]$). Beyond this limit, we again observe brittle fracture behaviour in tough-phase dominated two-phase solid
for ($r \geq 0.9$). Thus, the two-phase solids exhibit brittle fracture behaviour in the combined range of  ($r \in [0, 0.3] \cup [0.9, 1]$). The mechanisms responsible for the re-entrant brittle-ductile-brittle phase transition
will be discussed in the following sections with reference to their avalanche distribution and microstructure of damage.

\subsubsection{\label{sec:2.2}Avalanche distribution}
We will first study the avalanche distribution of the fragile and tough phases separately, followed by their combined effect on the avalanche distribution of the two-phase system. 

\begin{figure}
\centering
\begin{subfigure}[b]{0.45\textwidth}
\centering
\includegraphics[width=6.0cm, height=6.0cm]{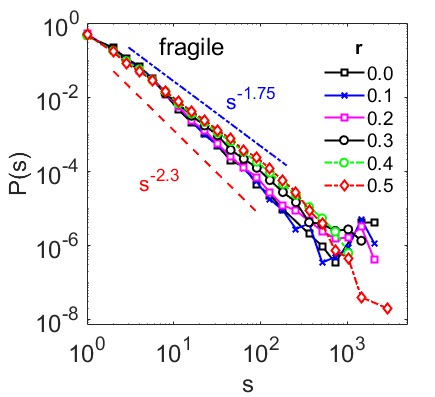}
\caption{}
\label{fig7a}
\end{subfigure}
\hfill
\begin{subfigure}[b]{0.45\textwidth}
\centering
\includegraphics[width=6.0cm, height=6.0cm]{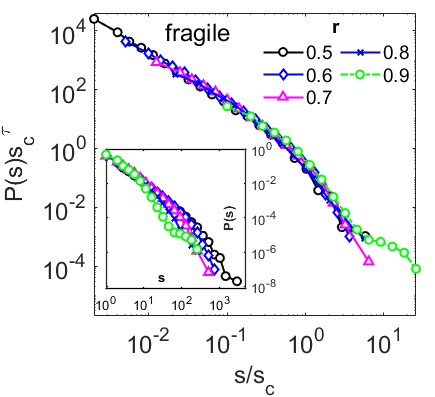}
\caption{}
\label{fig7b}
\end{subfigure}
\begin{subfigure}[b]{0.45\textwidth}
\centering
\includegraphics[width=6.0cm, height=6.0cm]{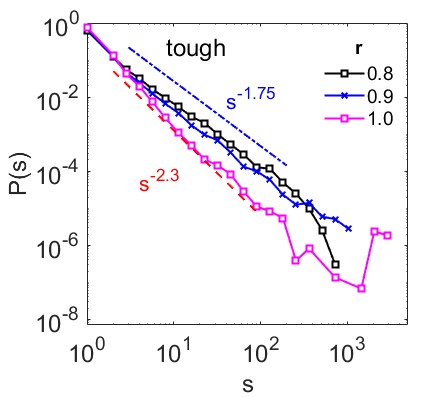}
\caption{}
\label{fig7c}
\end{subfigure}
\hfill
\begin{subfigure}[b]{0.45\textwidth}
\centering
\includegraphics[width=6.0cm, height=6.0cm]{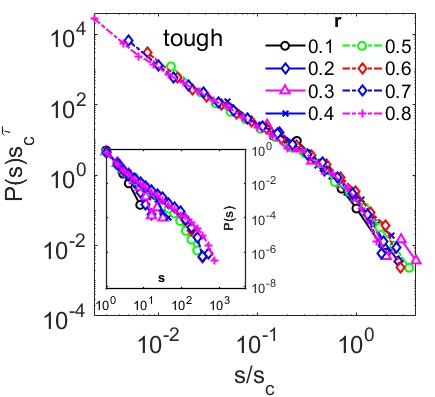}
\caption{}
\label{fig7d}
\end{subfigure}
\begin{subfigure}[b]{0.45\textwidth}
\centering
\includegraphics[width=6.0cm, height=6.0cm]
{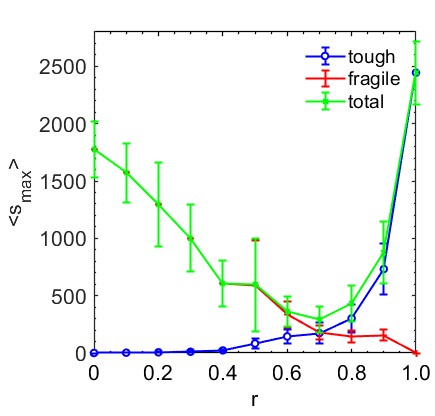}
\caption{}
\label{fig7e}
\end{subfigure}
\caption{For elastically matched solids ($\alpha = 1$, $\beta = 20$) with breakable tough phase, the avalanche distribution, for fragile phase, (a) $r \leq 0.5$, (b) $r \geq 0.5$ and for tough phase (c) $r \geq 0.8$, (b) $r \leq 0.8$. In (b) and (d), the rescaled avalanche distributions are shown in the main plot, with unscaled distributions in the inset. (e) The effect of $r$ on the maximum avalanche size $\langle s_{max} \rangle $.}
\label{fig7}
\end{figure}

The avalanche distribution for the homogeneous fragile phase ($r = 0.0$) exhibits a pure power-law behaviour with an exponent $\tau \approx 2.3$, as shown in Fig.~\ref{fig7a}. As the proportion of the fragile phase decreases, the power-law exponent of the avalanches of the fragile phase switches to approximately $1.75$, and an exponential cut-off appears in the tail due to system size effects. For $r > 0.3$, the nature of the power-law distribution remains unchanged, but the characteristic avalanche size $s_c$ decreases, as shown in Fig.~\ref{fig7b}. Interestingly, a switch is observed in the avalanche distribution of the tough phase also, when $r$ is decreased from $1$, as depicted in Figs.~\ref{fig7c} and \ref{fig7d}, where the exponent of the homogeneous tough phase is $2.3$. The switch in the exponent, however, occurs for the smallest fraction of fragile springs that we have considered. 
The switch in the exponent of the power-law distribution for the tough phase near $r=1$ occurs because damage in the tough phase initiates only after substantial failure has already occurred in the fragile phase distributed over the domain. The defect landscape created by prior failure of fragile phase permits damage in tough phase to grow in a more diffused manner, enabling larger but stable avalanches. These phase specific transitions are also reflected in the evolution of the maximum avalanche size, $\langle s_{max} \rangle $, within each phase. At $r=0$, $\langle s_{max} \rangle $ is largest in the fragile phase, consistent with brittle fracture where damage propagates catastrophically. As $r$ increases, the growing presence of tough springs constrains damage propagation in the fragile phase, thereby, reducing $\langle s_{max} \rangle $. In contrast, in the tough phase, a sharp transition occurs near $r =1$. At $r =1$ the tough phase has largest $\langle s_{max} \rangle $ implying brittle fracture. As fragile phase is introduced and $r$ decreases, $\langle s_{max} \rangle $ drops rapidly, indicating an abrupt shift toward more ductile-like fracture behaviours. Both the transitions are consistent with the changes in exponent observed in Fig.~\ref{fig7a} and Fig.~\ref{fig7c}.    

 While the individual avalanche distribution for fragile and tough springs show clear signatures of the transitions in fracture mechanisms, it is not clear what the implications are for the experimentally observable avalanche distribution which does not distinguish between fragile and tough springs. To understand the contribution of fragile and tough springs to the total avalanche distribution, we define the following quantity 
\begin{equation}
\label{eq10}
H(s) = \frac{\# \ of\  avalanche \ of \ size\  s}{\mathcal{N}},
\end{equation}
where, $\mathcal{N}$ is the same for total, fragile and tough springs. We choose $\mathcal{N}$ so that the total avalanche distribution is a probability density function. This definition allows us not only to compare the exponent but also the actual magnitudes of the contribution of fragile and tough avalanches.

\begin{figure}
\centering
\begin{subfigure}[b]{0.32\textwidth}
\centering
\includegraphics[width=4.5cm, height=4.5cm]{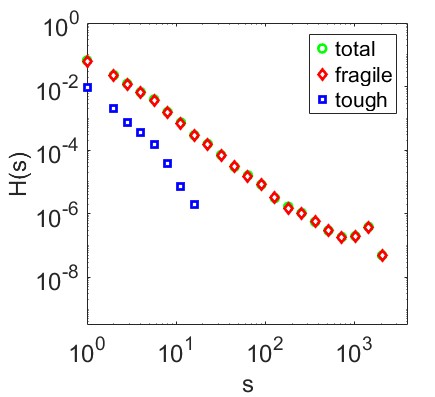}
\caption{$r$ = 0.2}
\label{fig8a}
\end{subfigure}
\begin{subfigure}[b]{0.32\textwidth}
\centering
\includegraphics[width=4.5cm, height=4.5cm]{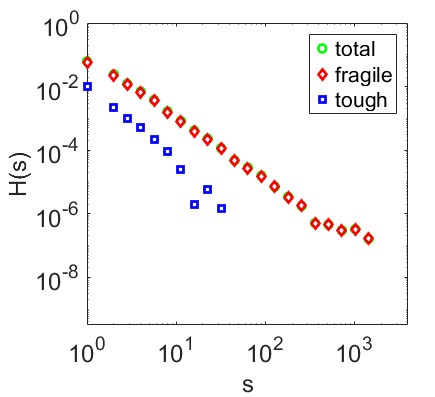}
\caption{$r$ = 0.3}
\label{fig8b}
\end{subfigure}
\begin{subfigure}[b]{0.32\textwidth}
\centering
\includegraphics[width=4.5cm, height=4.5cm]{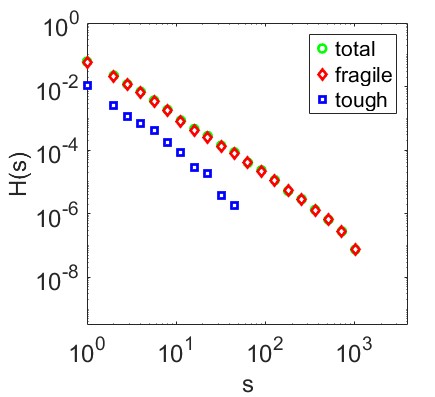}
\caption{$r$ = 0.4}
\label{fig8c}
\end{subfigure}
\begin{subfigure}[b]{0.32\textwidth}
\centering
\includegraphics[width=4.5cm, height=4.5cm]{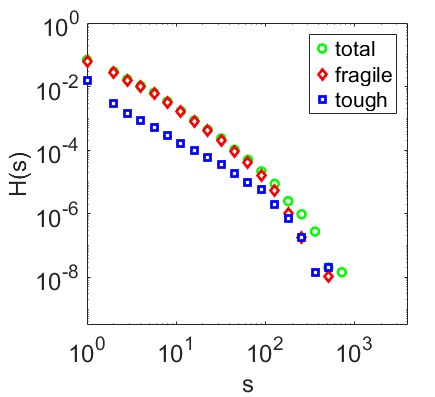}
\caption{$r$ = 0.7}
\label{fig8d}
\end{subfigure}
\begin{subfigure}[b]{0.32\textwidth}
\centering
\includegraphics[width=4.5cm, height=4.5cm]{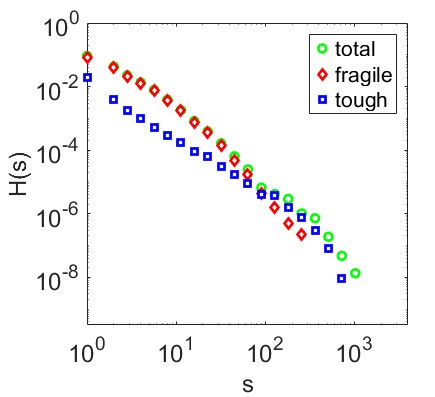}
\caption{$r$ = 0.8}
\label{fig8e}
\end{subfigure}
\begin{subfigure}[b]{0.32\textwidth}
\centering
\includegraphics[width=4.5cm, height=4.5cm]{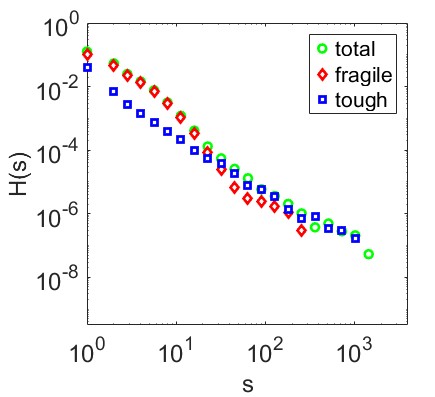}
\caption{$r$ = 0.9}
\label{fig8f}
\end{subfigure}
\caption{Avalanche distributions $H(s)$, (see Eq.$\sim$ \ref{eq10}) of the total number of broken springs along with that for fragile and tough springs are shown for different $r$. The data are for elastically matched solids ($\alpha = 1$, $\beta = 20$).}
\label{fig8}
\end{figure}

The total avalanche size distribution for small $s$ is dominated by the avalanche size distribution for fragile springs for all $r<1$ (see Fig.~\ref{fig8}). On the other hand, the avalanche distribution for large $s$ has contributions from both fragile and tough springs for $r\geq 0.7$ (see Figs.~\ref{fig8d}-\ref{fig8f}). For smaller $r$, the tough spring avalanches do not contribute much to the total avalanche distribution (see Figs.~\ref{fig8a}-\ref{fig8c}). While there are small avalanches of tough springs, their probability of occurrence is much lower. We also note that it becomes difficult to extract the exponent for the total avalanche distribution for larger $r$. Still the fracture mechanisms of a two-phase solid can be infered from the total avalanche distribution for large avalanches. A pure power-law behaviour observed for larger avalanches within the range ($r \in [0, 0.3] \cup [0.9, 1]$) signifies brittle fracture, while the presence of an exponential tail for other $r$ values indicates ductile-like fracture.

\subsubsection{\label{sec:2.3}Microstructure of damage}

\begin{figure}
\centering
\includegraphics[width=10.0cm, height=9.0cm]{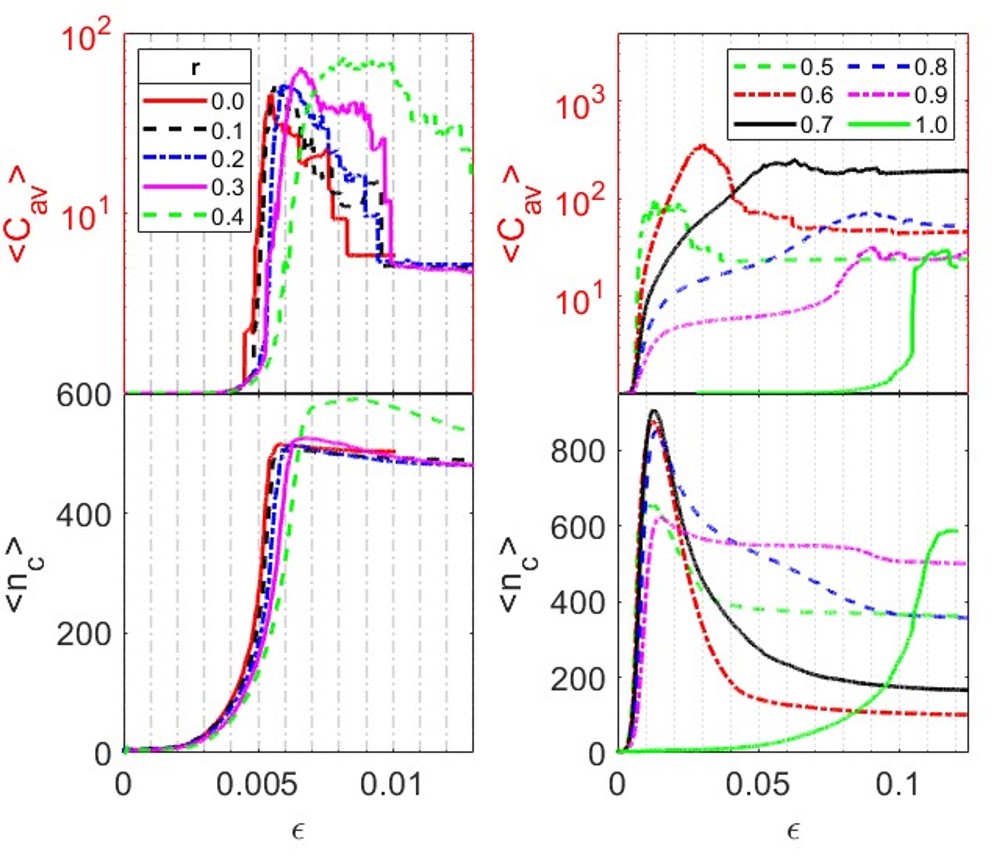}
\caption{The variation of average cluster size $\langle C_{av} \rangle $ of the broken springs and average number of clusters $\langle n_c \rangle $ with $\epsilon$ for various proportions breakable tough phase $r$ in the elastically matched solids ($\alpha = 1$. $\beta = 20$). Note that the largest cluster is omitted in the calculation of $\langle C_{av} \rangle $ for this figure.}
\label{fig9}
\end{figure}

The time evolution of the number $\langle n_c \rangle $ and size $\langle C_{av} \rangle $ of clusters (see Fig.~\ref{fig9}) for $r\leq 0.6$ closely resembles the behaviour observed for the two-phase solids having unbreakable tough phase, as shown in Fig.~\ref{fig5}. For larger $r$, the nature of the evolution in the microstructure of damage differs due to damage growth now possible in the breakable tough phase. Figure~\ref{fig9} reveals three distinct regimes of relative proportion, each exhibiting a unique time evolution of cluster characteristics: fragile phase dominated ($r \leq 0.3$), evenly mixed ($0.3<r<0.9$), tough phase dominated ($r\geq0.9$).

In the fragile phase dominant regime ($r \leq 0.3$), as shown in Fig~\ref{fig9}(a), clusters nucleate in the fragile phase and grow slowly up to a critical strain value. Beyond this point, $\langle C_{av} \rangle $ and $\langle n_c \rangle $ grow rapidly, suggesting accelerated growth of clusters as well as continued nucleation. They peak simultaneously at the same remote strain, which also corresponds to the peak in the macroscopic response. Further increase in applied strain results in drops in $\langle C_{av} \rangle $ with nearly stationary $\langle n_c \rangle $, suggesting rapid coalescence of a few larger clusters to form a single dominant cluster and failure. 

In the mixed regime ($0.3<r<0.9$), the initial clusters nucleate and grow in the fragile phase, until $\langle n_c \rangle $ reaches peak. The existing clusters then start coalescing in a stable manner restricted by the tougher phase such that $\langle C_{av} \rangle $ keeps increasing until it reaches its peak when rapid coalescence occurs as clusters merge through the tough phase into a single dominant cluster. The presence of the tougher phase, thus, allows the system to have a large number of clusters, and $\langle C_{av} \rangle $ for $r = 0.6$ is nearly four times that of a fragile phase-dominated regime. Beyond the peak, the dominant cluster absorbs other clusters, thus $\langle C_{av} \rangle $ decreases slowly and saturates as the damage activity slows down towards the final fracture. In this regime, after $\langle n_c \rangle $ reaches a peak, it exhibits a smooth exponential decline before saturating near the final fracture.  

In tough phase dominated two-phase solid ($0.9 \leq r< 1.0$), as shown in Fig.~\ref{fig9}(b), the time evolution of $\langle C_{av} \rangle $ follows the same pattern observed in the mixed regime. Similarly, the evolution of $\langle n_c \rangle $ remains unchanged until it reaches its peak. However, in the post-peak stage, instead of a smooth exponential decline up to the final failure, the rate of reduction suddenly increases at a critical strain value indicating rapid merging of clusters, and then it saturates towards the final failure. Notably, at this critical strain, $\langle C_{av} \rangle $ reaches its peak, which coincides with the peak in the macroscopic response. In the tough single-phase solid ($r = 1.0$), both $\langle C_{av} \rangle $ and $\langle n_c \rangle $ peak at the same strain, which coincides with the peak in the macroscopic response. Further loading shows a small reduction in $\langle C_{av} \rangle $ and $\langle n_c \rangle $ because after the peak macroscopic response, the solid catastrophically fractures followed by few damage activities.

\subsection{\label{sec:3} Combined effect of elastic heterogeneity and discontinuous threshold strain distribution}
In the previous section, we analysed the effect of discontinuity in threshold strain distribution between the constituent phases on the fracture behaviour of elastically matched solids. However, a more realistic scenario involves two-phase solids that exhibit both elastic mismatch and have differing failure threshold strains. To investigate the fracture behaviour of such two-phase solids, we consider a mixture of tough and fragile with ($\alpha = 20$ and $\beta = 20$) as shown in Fig.~\ref{figM2}(c), with the tough phase proportion denoted by $r$ as mentioned earlier in Sec.~\ref{sec: Mod2}. For clarity in depiction, we refer to the phases as hard-fragile and soft-tough to include the elastic characteristics when elastically mismatched.

\subsubsection{\label{sec:3.1}The constitutive behaviour}

\begin{figure}
\centering
\begin{subfigure}[b]{0.45\textwidth}
\centering
\includegraphics[width=6.0cm, height=5.8cm]{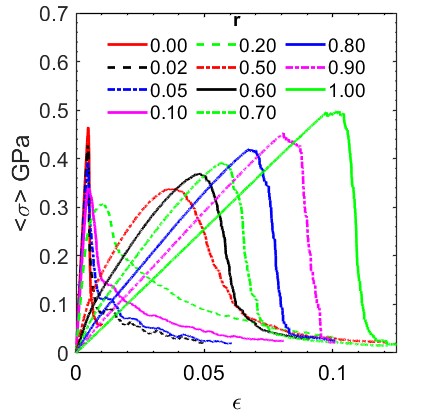}
\caption{}
\label{fig10a}
\end{subfigure}
\hfill
\begin{subfigure}[b]{0.45\textwidth}
\centering
\includegraphics[width=6.7cm, height=5.8cm]{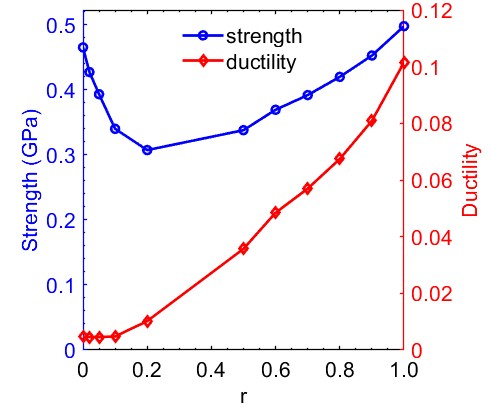}
\caption{}
\label{fig10b}
\end{subfigure}
\begin{subfigure}[b]{0.45\textwidth}
\centering
\includegraphics[width=6.3cm, height=6.3cm]{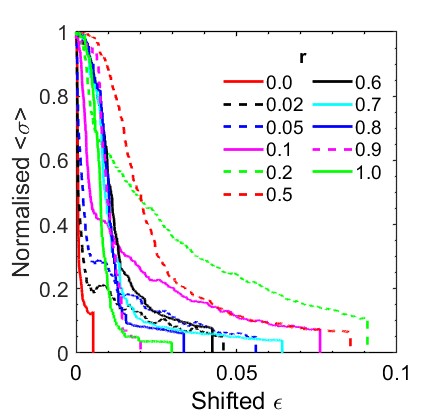}
\caption{}
\label{fig10c}
\end{subfigure}
\caption{(a) Constitutive behaviour of elastically mismatched solids ($\alpha = 20$, $\beta = 20$) for various fractions of the tough phase ($r$). (b) The effect of fraction of soft-tough phase on macroscopic strength ( $\langle \sigma \rangle_{max} $) and ductility ($\epsilon_{{\langle \sigma \rangle}_{max}}$). (c) The normalised post-peak stress-strain behaviour. For each $r$, $\langle \sigma \rangle $ is normalised by its corresponding peak, and $\epsilon$ is shifted by subtracting the strain value at $\langle \sigma \rangle_{max} $.}
\label{fig10}
\end{figure}

We first study the influence of the relative proportion of both constituent phases on the effective macroscopic response of the two-phase solid, presented in Fig.~\ref{fig10a}. The effective elastic modulus of the two-phase solid is, as expected, strongly dependent on the relative proportion of phases. As the proportion of the tough phase increases, the effective elastic modulus decreases from the maximum value of $E = 100$ GPa for the hard-fragile phase ($r = 0.0$) to the minimum value of $E = 5$ GPa for the soft-tough phase ($r = 1.0$), as shown in Fig.~\ref{fig10a}. The averaged stress-strain behaviour shows that the macroscopic strength, however, has a non-monotonic variation, as seen in Fig.~\ref{fig10b}, it is the highest for $r = 0$ and $1$  and then reduces as the degree of mixity increases in the two-phase solid, consistent with observation in~\cite{urabe2010fracture}. The underlying mechanisms resulting from the mismatch in modulus and failure strain threshold can be better understood by categorising the system response into three regimes. 

For low $r$, since the isolated regions of soft-tough phase have lower stiffness, they support reduced load, thereby, reducing the effective stiffness of the system. Even though the effective stiffness of system decreases, the ductility (strain at peak load) doesn't increase much as seen in Fig.~\ref{fig10b}, because the stress concentration in the hard-fragile phase triggers failure at lower load. Reduced stiffness and unaffected ductility result in reduced strength. 

For higher $r$, the presence of isolated hard-fragile phase is akin to that of inclusions in the soft-tough phase. The load supported by the hard-fragile phase is higher than the counterpart in a homogeneously soft-tough phase ($r = 1.0$). This load distribution increases the effective stiffness but also reduces the overall ductility. For instance, for $r = 0.8$, the reduction in ductility is $34\%$ while increase in stiffness is by $24\%$ leading to overall $16\%$ reduction in strength [see Fig.~\ref{fig10b}]. Notably, the reduction in strength for $r = 0.2$ (the reverse combination of phases) is more significant at $34\%$.

The normalised stress-strain behaviour beyond the peak is shown in Fig.~\ref{fig10c}. It reveals that the load drops sharply in two-phase solids with a smaller proportion of the tough phase ($r < 0.1$), showing brittle fracture behaviour. On further increase in the proportion of tough phase, the solid exhibits gradual load drop, showing ductile-like fracture behaviour in the range $r \in [0.1,0.6]$. Beyond this limit, we observe brittle fracture behaviour in soft-tough phase dominated solid for $r \geq  0.7$. Thus, the two-phase solids exhibit brittle fracture behaviour in the combined range of $r \in ([0.0,0.1) \cup [0.7,1.0])$. The mechanisms responsible for the brittle-ductile-brittle phase transition will be discussed in the following sections with reference to their avalanche distribution.    

\subsubsection{\label{sec:3.2}Avalanche distribution}

\begin{figure}
\centering
\begin{subfigure}[b]{0.45\textwidth}
\centering
\includegraphics[width=6.0cm, height=6.0cm]{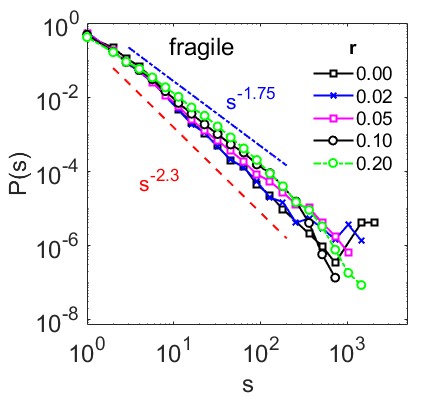}
\caption{}
\label{fig11a}
\end{subfigure}
\hfill
\begin{subfigure}[b]{0.45\textwidth}
\centering
\includegraphics[width=6.0cm, height=6.0cm]{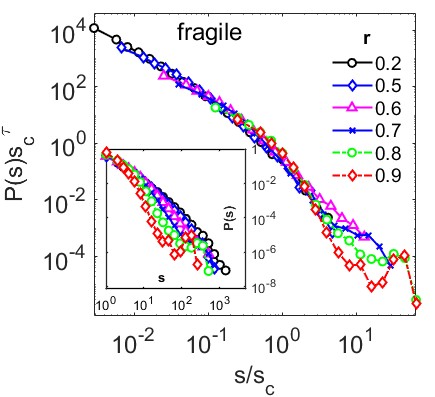}
\caption{}
\label{fig11b}
\end{subfigure}
\begin{subfigure}[b]{0.45\textwidth}
\centering
\includegraphics[width=6.0cm, height=6.0cm]{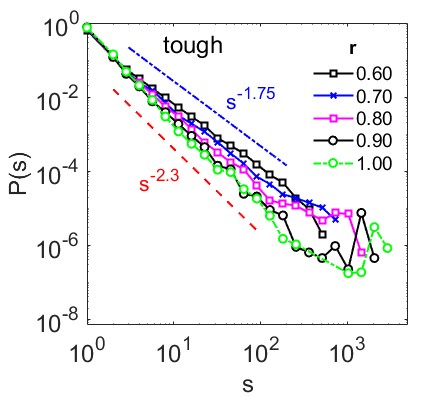}
\caption{}
\label{fig11c}
\end{subfigure}
\hfill
\begin{subfigure}[b]{0.45\textwidth}
\centering
\includegraphics[width=6.0cm, height=6.0cm]{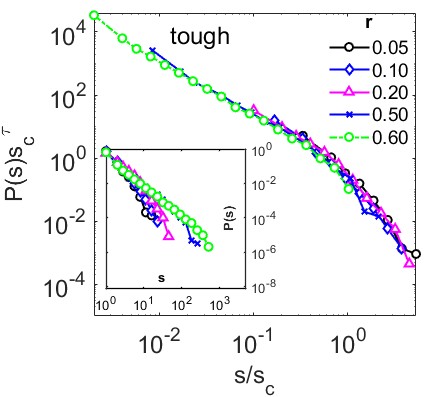}
\caption{}
\label{fig11d}
\end{subfigure}
\begin{subfigure}[b]{0.45\textwidth}
\centering
\includegraphics[width=6.0cm, height=6.0cm]
{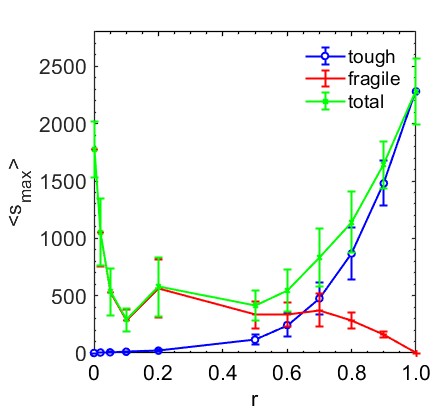}
\caption{}
\label{fig11e}
\end{subfigure}
\hfill
\begin{subfigure}[b]{0.45\textwidth}
\centering
\includegraphics[width=6.0cm, height=6.0cm]
{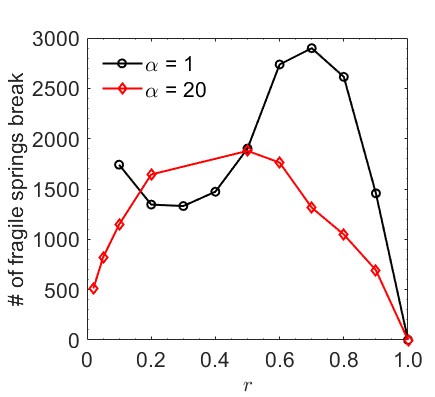}
\caption{}
\label{fig11f}
\end{subfigure}
\caption{Avalanche distribution in elastically mismatched solids ($\alpha = 20$, $\beta = 20$) for fragile phase, (a) $r \leq 0.2$, (b) $r \geq 0.2$ and for tough phase (c) $r \geq 0.6$, (b) $r \leq 0.6$. In (b) and (d), the rescaled avalanche distributions are shown in the main plot, with unscaled distributions in the inset. (e) The variation of the maximum avalanche size $\langle s_{max} \rangle $ with $r$. (f) shows the average number of fragile springs that fail before damage initiates in the tough phase for $\alpha = 1$ and $\alpha = 20$.}
\label{fig11}
\end{figure}

In this section, we first examine the avalanche distribution of the hard-fragile and soft-tough phases individually, followed by the total avalanche distribution of the two-phase system. The phase-wise avalanche distributions for the fragile and tough phases are presented in Fig.~\ref{fig11}. The avalanche distribution for both the phases demonstrate a pure power-law behaviour at higher proportions of the respective phase. However, it transitions to a power-law distribution with an exponential cut-off at lower proportions. This transition arises from the constraints imposed by one phase on the fracture dynamics of the other phase.

In the avalanche distribution of the hard-fragile phase shown in Fig.~\ref{fig11} (a) and (b) for the composition being predominantly fragile ($0 < r < 0.2$), a distinct switch of the exponent is observed in Fig.~\ref{fig11a}. The exponent switches from 2.3 to 1.75 for $r$ between 0.05 to 0.1. The switch is therefore, in the elastically mismatched case triggered at lower $r$ when compared to the elastically matched case earlier ($\alpha = 1$, $\beta = 20$), where the transition occurs for $r$ between 0.3 to 0.4. Also the hump at the tail of the distribution disappears with the switch in exponent, consistent with a single catastrophic event for lower $r$ and for $r$ in the range ($0.2 \leq r < 0.6$) more stable damage growth with increased probability of larger avalanches. For $0.2 < r < 0.9$, see Fig.~\ref{fig11b}, the distribution has the form of a power law with exponential cutoff with $s_c$ decreasing with increasing $r$. The hump at the tail reappears indicative of a single large event. Interestingly, the avalanche distribution of the soft-tough phase, see Fig.~\ref{fig11c}, also exhibits switch from $\tau \approx 2.3$ for homogeneous tough phase ($r = 1.0$). The switch occurs between $r = 0.8$ to $0.7$ i.e, $20-30\%$ of fragile phase. It is interesting to note that for both the phases, the switch is between the same exponents even though the fracture processes differ vastly. The presence of tougher phase in a predominantly fragile case increases resistance to rapid growth of damage due to a critical crack. In contrast, due to the presence of fragile phase in a predominantly tough phase, initial deformation results in breakage of nucleated fragile phase. The resulting distributed population of pores hinders the propagation along paths as in a corresponding homogeneous tough solid, thereby increasing the resistance to catastrophic growth of a critical crack. 

In Fig.~\ref{fig11e}, the largest avalanche of the hard-fragile is seen to drop rapidly, consistent with the switch in exponent for $r$ between 0.02 and 0.05. In the range $0.05<r<0.7$, the largest avalanche in both the phases is expectedly low. For compositions having $r>0.7$, the tough phase has a rapid increase, consistent with the reentrant transition from ductile-like to brittle-like fracture behaviour. Further insight into the mechanisms behind the switch in the exponent seen in both the phases, is gained by comparing the number of broken fragile phase prior to onset of any failure in the tougher phase at $\alpha = 1$, $\beta = 20$ and $\alpha = 20$, $\beta = 20$.  In Fig.~\ref{fig11f}, for $\alpha = 1$, the transition from brittle fracture for $r = 0$ to ductile-like fracture is due to the resistance to growth of nucleated cracks provided by the tough phase. Thus, the transition occurs only for a reasonable fraction of tough phase at $r \sim 0.3$. Expectedly the number of fragile springs that break before any damage initiation in the tough phase decreases with $r$ initially. Comparatively, for $\alpha = 20$, the transition is at a much lower fraction, $r\sim 0.05$. The difference is largely because of the stress concentration induced microcracking in fragile phase in the neighbourhood of the soft-tough phase. Any growth of micro-cracks involves interaction with soft-tough phase, thus, the number of fragile bonds that break is comparatively lower and increases with $r$ as the nucleation activity occurs around the soft-tough phase. The second transition for $\alpha = 1$ is for $r \sim 0.9$, as fragile phase, even a smaller fraction leaves behind a defect structure that arrests growth of micro-cracks resulting in stable ductile-like failure process. For $\alpha = 20$, however, the stiff inclusions build stress within and in neighbourhood. The failure events in the matrix and inclusions occur simultaneously. Growth of any defect initiated in the matrix has much less resistance in form of stressed inclusions rather than pores left behind after breakage of fragile bonds, as in $\alpha = 1$. Thus, the switch to ductile-like behaviour for fracture behaviour of tougher phase occurs at a higher fraction of the fragile phase for $\alpha = 20$. Unlike $\alpha = 1$, where the fragile bond breakage was significant before any breakage in the tough phase, the number of broken fragile springs is lower as the breakage in both phases are not as segregated.

\begin{figure}
\centering
\includegraphics[width=7.5cm, height=7.5cm]{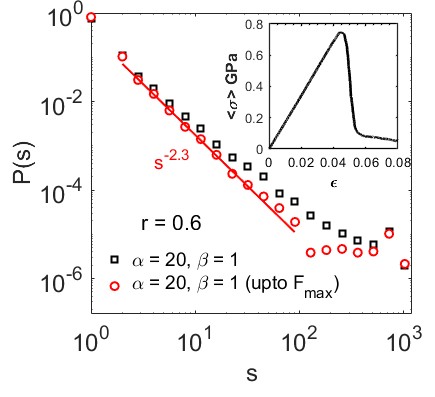}
\caption{ Avalanche distribution of the soft springs in solids that have only elastic mismatch
($\alpha = 1$, $\beta = 20$), for $r = 0.6$. Inset: The mean macroscopic stress-strain response.}
\label{fig12}
\end{figure}

Further, we have simulated another two-phase solid having $\alpha = 20$ with both phases having the same mean failure strain $\beta = \frac{\langle \epsilon_f \rangle _{soft}}{\langle \epsilon_f \rangle _{hard}} = 1$ and $r = 0.6$. The choice of properties ensures that early-stage damage does not occur in the hard phase alone and that both phases fail at similar strain. In Fig.~\ref{fig12}, the avalanche distribution of the soft springs exhibits a power-law behaviour for small avalanches with exponent $\approx 2.3$. As the avalanche size increases, the distribution deviates from the power-law. Additionally, the tail of the distribution displays a distinct hump, which arises from critical avalanches associated with brittle-like catastrophic failure, consistent with a sharp drop in the load seen in the macroscopic response after the peak load in the inset. When the avalanche series data beyond the peak load is omitted we see a clear power-law behaviour with exponent 2.3 extending even for larger avalanches. From this comparative avalanche distributions, one can infer that compared to the mismatch in failure strain threshold, the mismatch in elastic stiffness has less impact on the fracture statistics of the two-phase solid.

\begin{figure}
\centering
\begin{subfigure}[b]{0.32\textwidth}
\centering
\includegraphics[width=4.5cm, height=4.5cm]{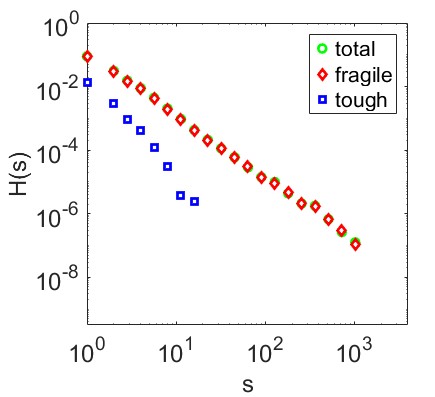}
\caption{$r$ = 0.05}
\label{fig13a}
\end{subfigure}
\begin{subfigure}[b]{0.32\textwidth}
\centering
\includegraphics[width=4.5cm, height=4.5cm]{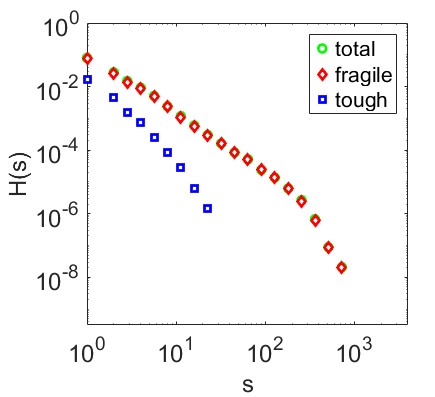}
\caption{$r$ = 0.1}
\label{fig13b}
\end{subfigure}
\begin{subfigure}[b]{0.32\textwidth}
\centering
\includegraphics[width=4.5cm, height=4.5cm]{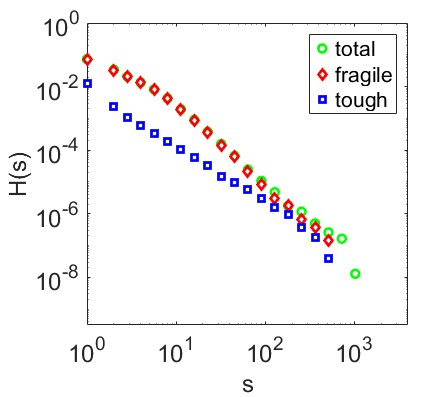}
\caption{$r$ = 0.6}
\label{fig13c}
\end{subfigure}
\begin{subfigure}[b]{0.32\textwidth}
\centering
\includegraphics[width=4.5cm, height=4.5cm]{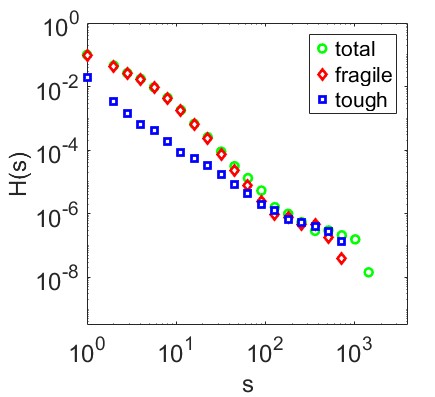}
\caption{$r$ = 0.7}
\label{fig13d}
\end{subfigure}
\begin{subfigure}[b]{0.32\textwidth}
\centering
\includegraphics[width=4.5cm, height=4.5cm]{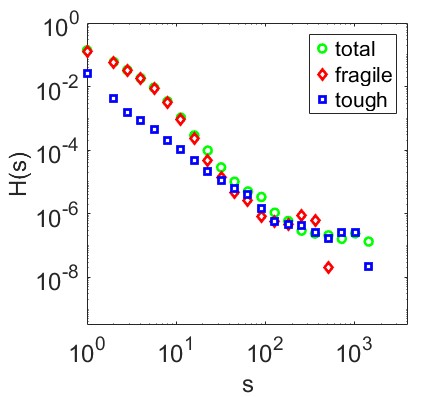}
\caption{$r$ = 0.8}
\label{fig13e}
\end{subfigure}
\begin{subfigure}[b]{0.32\textwidth}
\centering
\includegraphics[width=4.5cm, height=4.5cm]{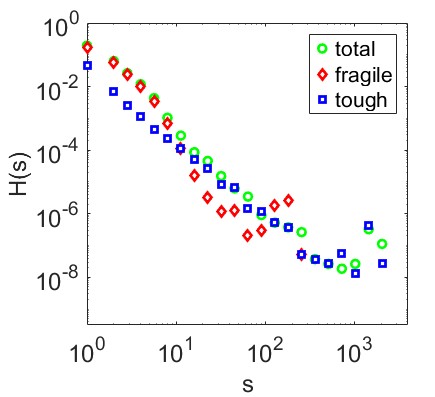}
\caption{$r$ = 0.9}
\label{fig13f}
\end{subfigure}
\caption{Avalanche distributions $H(s)$, (see Eq.$\sim$ \ref{eq10}) of the total number of broken springs along with that for fragile and tough springs are shown for different $r$. The data are for elastically mismatched solids ($\alpha = 20$, $\beta = 20$).}
\label{fig13}
\end{figure}

Understanding the avalanche dynamics of the constituent phases across various proportions, we next examine their influence on the overall avalanche dynamics of the system ($s_{total}$). The avalanche distribution of the total system, shown in Fig.~\ref{fig13}, shows that for lower relative proportion of tough phase, the $P(s)$ closely follows the statistics of the fragile phase showing a power-law with exponential cut-off. For the compositions with higher proportion of tough phase, the small avalanches continue to follow the statistics of the fragile phase while the larger avalanches tend to follow the tougher phase statistics. Thus, the total distribution in the range ($r>0.6$) has piecewise power-law statistics and the crossover between the regimes depend on $r$, higher the $r$ the crossover occurs at a small avalanche size. Further, as observed earlier for $\alpha = 1$ and $\beta = 20$, the total avalanche distribution, presented in Fig.~\ref{fig13}, for small $s$ is dominated by the contribution from the avalanches in the hard-fragile phase, except for $r=1$. For larger $s$, a crossover occurs where the total avalanches have the dominant contribution from the soft-tough phase. In the present case, the crossover occurs at smaller $s$ than in case of $\alpha = 1$ and $\beta = 20$. For instance for $r=0.8$, the crossover here is at $s\sim20$, (see Fig.~\ref{fig13e}), while earlier it was at $s\sim70$ (see Fig.~\ref{fig8e}).

\subsubsection{\label{sec:3.3}Microstructure of damage}

\begin{figure}
\centering
\includegraphics[width=10.0cm, height=9.0cm]{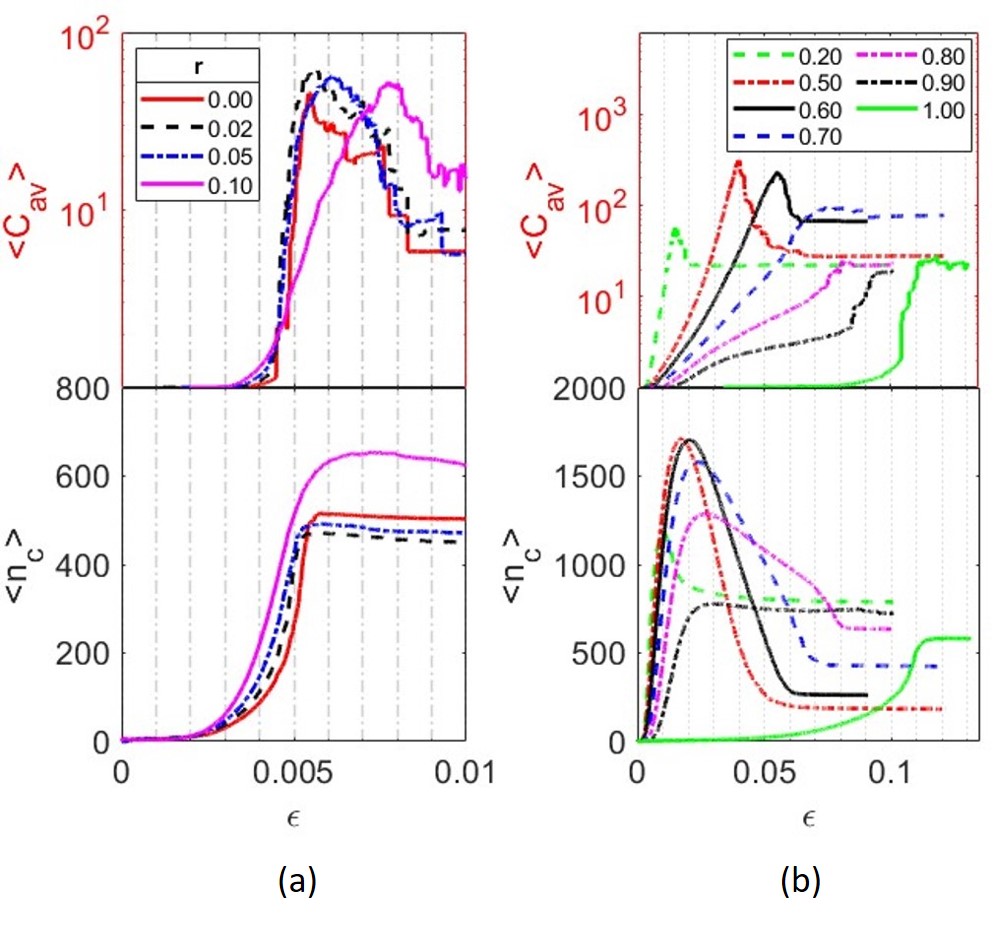}
\caption{The variation of average cluster size $\langle C_{av} \rangle $ of the broken springs and average number of clusters $\langle n_c \rangle $ with $\epsilon$ for various proportions breakable tough phase $r$ in the elastically mismatched solids ($\alpha = 20$, $\beta = 20$). Note that the largest cluster is omitted in the calculation of $\langle C_{av} \rangle $ for this figure.}
\label{fig14}
\end{figure}

In the cluster statistics of the elastically mismatched system ($\alpha =20$, $\beta =20$), we see again three distinct regimes: fragile phase dominated ($0.0<r<0.1$), heterogeneous ($0.2<r<0.6$), tough phase dominated ($r\geq0.7$). In ($0.0<r<0.1$) regime, clusters nucleate in the fragile phase and grow slowly until $\epsilon \sim (\epsilon_f)_{fragile}$ when $\langle C_{av} \rangle $ and $\langle n_c \rangle $ grow rapidly. Both parameters peak simultaneously. Further increase in applied strain results in rapid drop in $\langle C_{av} \rangle $ with nearly stationary $\langle n_c \rangle $ suggesting rapid coalescence of a few larger clusters to form a single dominant cluster and failure. This regime notably has over all lesser clusters and $\langle C_{av} \rangle $ never exceeds 50. In the next regime ($0.2<r<0.6$), the initial clusters nucleate and grow in the fragile phase, until $\langle n_c \rangle $ reaches peak. The existing clusters then start coalescing in a stable manner restricted by the tougher phase such that $\langle C_{av} \rangle $ keeps increasing until it reaches its peak when rapid coalescence occurs as clusters merge through the tough phase into a single dominant cluster. The presence of tougher phase, thus, allows the system to have more than three times number of clusters and $\langle C_{av} \rangle $ for $r = 0.5$ is nearly four times that of fragile phase dominated regime, implying void spread diffused damage. The tough phase dominated regime has the usual initial nucleation and growth in the fragile phase, followed by coalescence. The resulting growth in cluster size is slow. As the remotely applied strain reaches near critical conditions for growth in tougher phase $\langle C_{av} \rangle $ increases rapidly as the clusters grow into the tougher phase and the solid fracture. In this regime, like the fragile phase-dominated regime, the overall number of clusters and the largest observed clusters are significantly lower than in the heterogeneous regime, implying brittle fracture.

\section{\label{sec:4}Conclusions}
The complex processes during fracture of multi-phase solids inherently involve several competing mechanisms with many contributing factors.

In the present study, using random spring network model,  we investigate the specific role of elastic mismatch and failure strain mismatch on the macroscopic fracture response of a two-phase solid, focusing on the type of fracture and the underlying mechanisms. Although we take each phase to be individually brittle, when combined, the two-phase solid is shown to exhibit a transition to ductile-like behavior over a specific range of volume fractions.

At the macroscopic scale, the avalanche size distribution of the two-phase composite does not follow a clear power-law. However, when analyzed separately, the avalanche distributions of the individual phases, consistently exhibit power-law behavior across all mixing ratios. Notably, the brittle-to-ductile transition at small fractions of the tough phase in the two-phase solid is marked by a change in the power-law exponent of the avalanche distribution of the fragile phase, whereas the reentrant ductile-to-brittle transition at large fractions of the tough phase is characterized by a corresponding shift in the avalanche statistics of the tougher phase. 

Interestingly, the critical fraction of the minority phase at which these transitions occur differs between the two cases and is strongly influenced by the elastic and fracture property contrasts between the phases. The observed transitions in fracture behaviour are primarily driven by the mismatch in failure strain thresholds between the constituent phases. When the fragile phase is in the majority, toughening arises due to the constraint on crack growth imposed by the embedded tough phase. In contrast, when the tough phase dominates, the toughening mechanism shifts: it is now governed by the defect structure left behind after the initial failure of the fragile phase. This defect structure facilitates damage growth through stable fracture events between interacting defects, which release strain energy gradually and as a consequence, inhibit catastrophic crack propagation.

Elastic mismatch further modifies these mechanisms. When the hard-fragile phase is in the majority, additional toughening occurs because stress concentrates near the soft-tough phase, promoting nucleation near the interface. As a result, the influence of the soft-tough phase is significant even when it is present in smaller fractions. Conversely, when the soft-tough phase is in the majority, the elastic mismatch suppresses toughening mechanisms. Stress builds up in both phases, leading to simultaneous nucleation in the soft-tough and hard-fragile phases. The micro-cracks initiated in the soft-tough phase grow rapidly and face little resistance, as the minority hard-fragile phase offers minimal constraint.

The differences in the way damage propagates in the brittle and ductile regimes were captured by the time evolution of cluster characteristics of broken springs. In the brittle regime, the dominant cluster grows rapidly absorbing other large clusters, leaving number of clusters roughly unchanged. In the ductile regime, coalescence is the more dominant process, resulting in decrease in the total number of clusters with time while the mean cluster size grows.

\section{\label{sec:5} Discussion}

We now contrast our results with earlier studies that had simulated fracture in two-phase solids using statistical models. Within the mean field-like fibre bundle model (FBM), the avalanche statistics have been studied for global load sharing as well as the local load sharing model. In the global load-sharing model, when only one phase is breakable, a phase transition occurs when the fraction of the unbreakable springs is increased. The avalanche exponent switches from $5/2$ to $9/4$ with an exponential cutoff for larger fractions of unbreakable bonds~\cite{hidalgo2008universality}. When the model was studied with local load sharing rules, then the transition occurs at a very small fraction of the unbreakable springs, with the exponent shown numerically to shift from 4.5 to 2.0~\cite{kovacs2013brittle}. The global load sharing model has also been studied with strain thresholds that have a discontinuity in the threshold strength distribution. The discontinuity leads to a non-universal behaviour in the avalanche size distribution for smaller values of avalanche size~\cite{divakaran2007effect}. Fracture in the two-phase solid has also been studied in a central force-based RSNM comprising breakable and unbreakable phases~\cite{noguchi2024fracture}. Unlike the fibre bundle model, no transitions were found, and the avalanche exponent remains unchanged with introduction of unbreakable bonds. We note that the unbreakable springs were positioned only along the boundaries of square unit cells that enclose the breakable springs. 

The first transition from brittle to ductile fracture, found in this paper, is similar to that found for fibre bundle model with global load sharing and one of the phases being unbreakable. Unlike the fibre bundle model, the elastic continuum behaviour is well captured by the RSNM, and hence the corresponding exponents are expectedly different. In our simulations, the avalanche exponent switches from $\approx 2.33$ to $\approx 1.75$. We also find a second ductile to brittle transition as the fraction of tough bonds is increased. This was not found for fibre bundle model with global load sharing. We point out that in the latter case, the threshold discontinuity and fraction of tough bonds were not varied independently~\cite{divakaran2007effect}. Compared to the fracture simulations using RSNM~\cite{noguchi2024fracture}, we obtain a rich and varied macroscopic response. We attribute this difference to the fewer cases and narrow range of the relative fractions of fragile and unbreakable springs that were studied in~\cite{noguchi2024fracture}, as well as the correlated architecture of unbreakable bonds that was used.

We now discuss how the existence of two avalanche exponents in a two-phase solid, depending on the relative fraction of the tough and fragile phases, allows us to reach a better understanding of the universality classes of avalanche exponents during fracture. There have been many studies of damage growth in materials undergoing fracture through the analysis of the accompanying acoustic emissions. The exponents for the power law distributed acoustic emission data in these controlled “lab-quakes” have been argued to fall into broadly two universality classes~\cite{xu2019criticality}. One where the exponent is $\approx 1.67$ and the other where the exponent is $\approx 1.4$. While some characteristic features of the materials in the two universality classes have been contrasted~\cite{xu2019criticality}, there is no well-established understanding. Also, there is no clear understanding within models like RSNM or random fuse models for the different universality classes. The avalanche exponent in these statistical models is related to the exponents of acoustic emission as $\tau_{AE} = (1 + \tau_s)/2$, where $\tau_{AE}$ is the acoustic emission exponent and $\tau_s$ is the spring or fuse avalanche exponent. Thus, the experimentally observed universality classes would correspond to $\tau_s \approx 2.33$ and $\tau_s \approx 1.8$. The former is found in generic simulations of RSNM~\cite{senapati2024acoustic,kumar2022interplay,ray2006breakdown,zapperi1997first}. The exponent $1.8$ was found for RSNM simulations of bone, where the disorder was taken from the CT scan images of bone~\cite{mayya2017role,mayya2018role}. However, the precise reason for the shift was not understood.

\begin{figure}
\centering
\includegraphics[width=7.0cm, height=7.0cm]{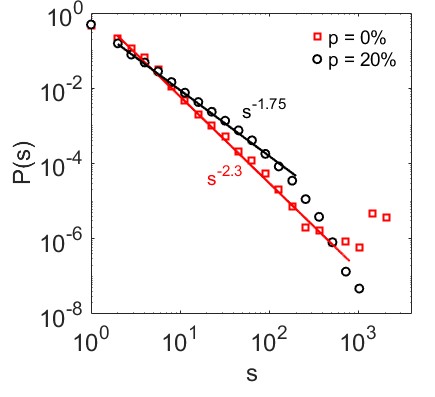}
\caption{The avalanche distribution in a single-phase solid for no porosity and $20\%$ porosity.}
\label{fig15}
\end{figure}

 In this paper, we found that when a small fraction of fragile bonds are added to a matrix of tough bonds, the avalanche exponent switches to $1.75$. We also observe that when strain is applied, then it is the fragile bonds that break first. Given this, we conjecture that if we take a single-phase RSNM and increase porosity, then the exponent should switch from $2.3$ to $1.75$, an increase in porosity mimicking the inclusion of a fragile phase in a tough material. In Fig.~\ref{fig15}, we show results of simulations of RSNM for a homogeneous system with no porosity and $20\%$ porosity, obtained by breaking the springs attached to $20\%$ of randomly chosen sites. It is clear that the exponent shifts from $\approx 2.3$ to $\approx 1.75$, providing an explanation for the universality classes as seen in experiments of fracture. A more detailed study of this transition with simulations of larger system sizes is a promising area of future study.

\bibliography{aipsamp}% Produces the bibliography via BibTeX.

\end{document}